\documentclass[final,3p,times]{elsarticle}

\usepackage{amssymb}

\usepackage{amssymb}
\usepackage[]{amsmath}
\usepackage{graphics}
\usepackage{amssymb}
\usepackage[]{amsmath}
\usepackage{amsthm}
\usepackage{setspace}
\usepackage{epsfig}
\usepackage{subfigure}
\usepackage{cases}
\usepackage{graphicx}

\biboptions{numbers,sort&compress}

\usepackage{amsmath}
\usepackage{times}
\usepackage{anysize}
\marginsize{2.5cm}{2.5cm}{1cm}{2cm}

\selectfont

 \journal {}

\begin{document}

\begin{frontmatter}



\title{Dynamics of Rogue Waves in the Partially $\cal{PT}$-symmetric\\
 Nonlocal Davey-Stewartson Systems}

\author{ Bo Yang \fnref{label1,label2} }
\author{ Yong Chen \fnref{label1,label2} \corref{cor1} }
\ead{ychen@sei.ecnu.edu.cn.}

\cortext[cor1]{Corresponding author}

\address[label1]{Shanghai Key Laboratory of Trustworthy Computing, East China Normal University, Shanghai, 200062, People's Republic of China}

\address[label2]{MOE International Joint Lab of Trustworthy Software, East China Normal University, Shanghai, 200062, People's Republic of China}
\begin{abstract}
 In this work,  we  study the dynamics of rogue waves in the partially $\cal{PT}$-symmetric nonlocal  Davey-Stewartson(DS) systems. Using the Darboux transformation method, general rogue waves in the partially $\cal{PT}$-symmetric nonlocal DS equations are derived. For the partially $\cal{PT}$-symmetric nonlocal DS-I equation, the solutions  are obtained  and expressed  in term of  determinants.  For the partially $\cal{PT}$-symmetric DS-II equation,  the solutions are represented as quasi-Gram determinants.   It is shown that the fundamental rogue waves in these two systems are rational solutions which arises from a constant background at $t\rightarrow -\infty$, and develops finite-time singularity on an entire hyperbola in the spatial plane at  the critical time.  It is also shown that the interaction of several fundamental rogue waves is described by the  multi rogue waves. And the interaction of fundamental rogue waves with dark and anti-dark rational travelling waves generates the novel hybrid-pattern waves.  However, no high-order rogue waves are found in this partially $\cal{PT}$-symmetric nonlocal DS systems. Instead, it can produce some high-order travelling waves from the high-order rational solutions.
\end{abstract}

\begin{keyword}
Nonlocal Davey-Stewartson equations,\  Darboux transformation, \ Rogue waves

\end{keyword}
\end{frontmatter}



\section{Introduction}
It is known to us all that the integrable nonlinear evolution equations are exactly solvable models which play an important role in a lot of branches of nonlinear science, especially in the study of nonlinear physical systems, including water waves, nonlinear optics, Bose-Einstein condensates  and plasma physics. There are  numerous celebrated continuous and discrete integrable systems that are physically revelent.   In particular, the  nonlinear Schr\"{o}dinger (NLS)~\cite{Faddeev1987} and the Davey-Stewartson (DS)~\cite{Davey1974}
equations  are classical examples of generic integrable PDEs. The NLS-type equations are the essential models describing optical wave propagation in nonlinear optics. The DS equations, which can be seen as the multi-dimensional extension of NLS equation, are also the universal models governing the evolution of two-dimensional wave packet on water of finite depth.

In the last several years, $\mathcal{PT}$-symmetric systems, which allow for lossless-like propagation due to their balance of gain and loss, have attracted considerable attention and triggered renewed interest in integrable systems.  Quite a lot of work were done on the new nonlocal integrable systems \cite{AblowitzMussPRL2013,AblowitzMussPRE2014,Yan,Khara2015,AblowitzMussNonli2016,Zhu1,Fokas2016,Lou,Lou2,AblowitzMussSAPM,Chow,ZhoudNLS,ZhouDS,HePTDS,HePPTDS,Zhu2,Zhu3,Gerdjikov2017,
Ablowitz_arxiv, BoTransformations}. These nonlocal integrable equations are different from local integrable equations and could produce  novel patterns of solution dynamics and intrigue new physical applications. Among these models, the $\mathcal{PT}$-symmetric  NLS equation was the first nonlocal integrable equation  proposed in \cite{AblowitzMussPRL2013}:
\begin{eqnarray}\label{PTNLS}
  iq_{t}(x,t)=q_{xx}(x,t)+V(x,t)q(x,t),
\end{eqnarray}
with $ V(x,t)=-2\sigma q(x,t)q^{*}(-x,t),\ \sigma=\pm1$. It is shown to be an integrable infinite dimensional Hamiltonian equation with a self-induced potential  satisfying the $\cal{PT}$-symmetry condition: $V(x,t)=V^{*}(-x,t)$. The nonlocality  occurs in the form that one of the nonlinear terms is dependent on variable evaluated at $-x$.   One-soliton solution with singularity for the focusing nonloc-NLS equation (\ref{PTNLS}) has been obtained via the inverse scattering transform (IST). More detailed study of the inverse scattering theory for eq.(\ref{PTNLS}) was developed and the Cauchy problem was formulated  in  \cite{AblowitzMussNonli2016} via the Riemann-Hilbert problem (RHP).

As an integrable multidimensional versions of the nonlocal nonlinear Schr\"{o}dinger equation, a new integrable nonlocal Davey-Stewartson (DS) equation is recently introduced in Refs.\cite{Fokas2016, AblowitzMussSAPM}:
\begin{eqnarray}\label{PTDS}
  && iu_{t}+\frac{1}{2}\alpha^{2} u_{xx}+\frac{1}{2} u_{yy}+ (u v - \emph{w})u=0, \nonumber \\
  && w_{xx}-\alpha^{2} w_{yy} - 2\left[(u v)\right]_{xx}=0,
\end{eqnarray}
 where $u$, $v$ and $w$ are functions of $x,y,t$, $\alpha^2=\pm1$ is the equation-type parameter(with $\alpha^2=1$ being the DS-I and $\alpha^2=-1$ being DS-II).  With different symmetry reductions of potential function $u$ and $v$, this equation contains two nonlocal versions:\\
(i).  \ $\mathcal{PT}-$ symmetric nonlocal reduction : $v(x,y,t)=\epsilon \bar{u}(-x, -y,t)$;  \\
(ii).  \ Partially\  $\mathcal{PT}-$ symmetric nonlocal reduction :  $v(x,y,t)=\epsilon \bar{u}(-x,y,t)$,\\
here the sign $\bar{u}$ represents the complex conjugation of this function, and $\epsilon=\pm1$ is the sign of nonlinearity. For these two nolocal versions, several results have been obtained in \cite{ZhouDS, HePTDS,HePPTDS} by Darboux transformation or the Hirota bilinear method. Furthermore, other versions of nonlocal DS equations are also proposed and studied in \cite{AblowitzMussSAPM} according to different types of time-space coupling. Especially, when $v(x,y,t)=\epsilon \bar{u}(x,-y,t)$, it produces another version of partially $\cal{PT}$-symmetric nonlocal DS equations.

In this article, we focus on the nonlocal DS systems with  partially $\cal{PT}$-symmetric potential (i.e., the nonlocal version (ii)). Using the Darboux transformation method, general rogue waves in the partially $\cal{PT}$-symmetric nonlocal DS equations are derived.  On the one hand, solutions of the partially $\cal{PT}$-symmetric nonlocal DS-I equation   are obtained  and expressed  in terms of  determinants.  On the other hand,  through the binary DT, solutions of the partially $\cal{PT}$-symmetric DS-II equation are constructed and represented as quasi-Gram determinants.

With different parameters chosen in the fundamental rational solutions,  it is shown that the fundamental rogue waves in these two systems are rational solutions which arises from a constant background at $t\rightarrow -\infty$, and develops finite-time singularity on an entire hyperbola in the spatial plane at  the critical time.  It is also shown that the interaction of several fundamental rogue waves is described by the  multi rogue waves, which are generated from multi-rational solutions,  and the singular time points in these   multi rogue waves  appear in pairs or in a time interval.
It is further shown that the interaction of fundamental rogue waves with dark and anti-dark rational travelling waves generates the hybrid-pattern waves. This novel pattern, which contains three different wave patterns in one solution,  to the best of our knowledge, has never been reported in the local and nonlocal DS systems.

As we know, the local DS systems possess several patterns of high-order rogue waves~\cite{YangDSI,YangDSII}. However, in this partially $\cal{PT}$-symmetric nonlocal DS systems, we can not find any high-order rogue waves, even though there are  some high-order rational travelling waves produced from the high-order rational solutions.  This might because the singularity in the fundamental rogue waves  will quickly increase if one proceeds  the iteration  through  the high-order DT. While the iteration of $N$-fold DT only increase the numbers or the range of singularities, as what we have shown in the multi-rogue waves.
It is found in  refs.~\cite{YangJOL2014,KKTLPRL2015,JYPRE2015,BKBPRA20152015,KJYADRMP2016} that some possible applications in optics  have  been shown  in the partially $\cal{PT}$-symmetric physical systems.  We expect these rogue-wave solutions could have interesting implications for  partially $\cal{PT}$-symmetric   in multi-dimensions.

\section{Darboux transformation for nonlocal DS system}
In this section,  we first work on the form of Darboux transformation in the general Davey-Stewartson system with the partially  $\cal{PT}-$symmetric nonlocal reduction. For eqs.(\ref{PTDS}), the corresponding  auxiliary linear system  is reduced from the (2+1) dimensional AKNS system:
\begin{eqnarray}
&&\Phi_{y}=J\Phi_{x}+P \Phi,  \label{2+1AKNS1} \\
&&\Phi_{t}=\sum_{j=0}^{n} V_{n-j} \partial^{j} \Phi, \label{2+1AKNS2}
\end{eqnarray}
where $\partial=\partial/\partial x$, $V_{j}$ are $N \times N$ matrices, $J$ is   $N \times N$ constant diagonal matrix, and $P$ is a $N \times N$ off-diagonal   matrix.

Taking $N=2$, $n=2$ in (\ref{2+1AKNS1})-(\ref{2+1AKNS2}), it  generate the following Lax-pair for systems (\ref{PTDS}) :
\begin{eqnarray}
&&\emph{L}\Phi=0,\ \  \emph{L}=\partial_{y}-J\partial_{x}-P, \label{Lax-pair1} \\
&& \emph{M}\Phi=0,\ \ \emph{M}=\partial_{t}-\sum_{j=0}^{2}V_{2-j}\partial^{j}=\partial_{t}-i\alpha^{-1} J \partial_{x}^{2}-i\alpha^{-1} P \partial_{x} - \alpha^{-1} V,\label{Lax-pair2}
\end{eqnarray}
where,
\begin{eqnarray*}
  && J=\alpha^{-1}\left(
                    \begin{array}{cc}
                      1 & 0\\
                      0 & -1 \\
                    \end{array}
                  \right),\
  P=              \left(
                    \begin{array}{cc}
                      0 & u\\
                      -v & 0 \\
                    \end{array}
                  \right),
   \\
  && V= \frac{i}{2}\left(
                    \begin{array}{cc}
                      \omega_{1} & u_{x}+\alpha u_{y}\\
                      -v_{x}+\alpha v_{y} & \omega_{2}\\
                    \end{array}
                  \right),
\end{eqnarray*}
with
\begin{eqnarray}
 w=uv-\frac{1}{2 \alpha}(\omega_{1}-\omega_{2}).
\end{eqnarray}
With the partially\  $\mathcal{PT}-$ symmetric reduction $v(x,y,t)=\epsilon \bar{u}(-x,y,t)$, the integrability condition:
\begin{center}
$\Phi_{y,t}=\Phi_{t,y}$
\end{center}
leads to the  partially\  $\mathcal{PT}-$ symmetric nonlocal DS equations.

\subsection{Darboux transformation for  partially $\cal{PT}$-symmetric  nonlocal DS-I}
It is already  known in \cite{MatveevDT1991, NimmoGY2000} that for any invertible matrix $\theta$ such that $L(\theta)=M(\theta)=0$, the operator
\begin{eqnarray}\label{eDT}
G_{\theta}=\theta \partial \theta^{-1},\  \ \partial=\partial_{x},
\end{eqnarray}
 makes $L$ and $M$ form invariant under the elementary Darboux transformation:
\begin{eqnarray*}
L\rightarrow \tilde{L}=G_{\theta}  L  G_{\theta}^{-1},\ M\rightarrow \tilde{M}=G_{\theta}  M  G_{\theta}^{-1}.
\end{eqnarray*}

Next,  we introduce some notations. For  operator $L$ and its adjoint operator $L^{\dag}$, defining  the space $S$ and $S^{\dag}$ which  stand for the sets of nontrivial solutions in the kernel of the operator, i.e.,:
\begin{eqnarray*}
&&S=\{\theta, \ \theta \ \text{is nonsingular}: L(\theta)=0\},\\
&&S^{\dag}=\{\rho, \ \rho \ \text{is nonsingular}: L^{\dag}(\rho)=0\},
\end{eqnarray*}
and define $\tilde{S}$, $\tilde{S}^{\dag}$ for operator $\tilde{L}$, $\tilde{L}^{\dag}$ etc.
Thus, this elementary DT (\ref{eDT}) defines the mapping:
\begin{center}
$G_{\theta}: S\rightarrow \tilde{S}$.
\end{center}

For the  Darboux transformation, as we known, if we pose some restriction to the potential (e.g. this $v(x,y,t)=\epsilon \bar{u}(-x, -y,t)$ in the  partially $\cal{PT}$-symmetric  nonlocal DS equations), then the transforation does not naturally preserve the conditions.  Therefore, in this case, we need more restrictions on the choices of solution matrix $\theta$.

Let
$
\sigma=
\left(
  \begin{array}{cc}
    0 & -\epsilon  \\
    1 & 0 \\
  \end{array}
\right)
$ ,  then  potential matrix  $P$ and $V_{2}$ in (\ref{Lax-pair1})-(\ref{Lax-pair2}) satisfy the following symmetric reduction
\begin{eqnarray}\label{SPreduction}
\sigma P(x,y,t) \sigma^{-1}=\overline{P}(-x,y,t),\  \sigma V_{2}(x,y,t) \sigma^{-1}=\overline{V}_{2}(-x,y,t),
\end{eqnarray}
here we need the property $\omega_{1}(x,y,t)=-\overline{\omega_{2}}(-x,y,t)$,  which can be derived form the integrability condition.

This give rise to the symmetry constraint in $L, M$:
\begin{eqnarray}\label{SLreduction}
\sigma L \sigma^{-1}=\overline{L}_{(x\rightarrow-x)},\ \sigma M \sigma^{-1}=\overline{M}_{(x\rightarrow-x)}.
\end{eqnarray}
Suppose
$\left(
\begin{array}{c}
\xi(x,y,t)\\
\eta(x,y,t)
\end{array}
\right) $
is a vector solution of equations (\ref{Lax-pair1})-(\ref{Lax-pair2}), it is inferred  from symmetry (\ref{SLreduction}) that
$
\left(
\begin{array}{c}
-\epsilon \bar{\eta}(-x,y,t)\\
\bar{\xi}(-x,y,t)
\end{array}
\right)
$
is also a solution. Hence we can choose the matrix $\theta$ as:
\begin{eqnarray}\label{solutionmatrix}
\theta=
\left(
  \begin{array}{cc}
    \xi(x,y,t) & -\epsilon \bar{\eta}(-x,y,t)  \\
    \eta(x,y,t) & \bar{\xi}(-x,y,t) \\
  \end{array}
\right),
\end{eqnarray}
and  $\theta$ also admits the symmetry
\begin{eqnarray}\label{WFLreduction}
 \overline{\theta}(x,y,t) = \sigma \theta(-x,y,t) \sigma^{-1}.
\end{eqnarray}

Since the n-fold   DT is nothing but a n-times iteration of the one-fold DT, we merely consider the one-fold DT.
With the action of elementary DT, we obtain the relation between potential matrices:
\begin{eqnarray}\label{Npotential}
 && \widetilde{P}=P+ [J, S],\ S=\theta_{x}\theta^{-1}, \\
 && \widetilde{V}_{2}=V_{2}+V_{1,x}+2V_{0}S_{x}+[V_{0}, S]S+[V_{1}, S].
\end{eqnarray}

Moreover, it can be verified that transformation $G_{\theta}$  keep the reduction relation (\ref{SPreduction}) and (\ref{SLreduction}) invariant, i.e:
\begin{eqnarray*}
&&\sigma \widetilde{P}(x,y,t) \sigma^{-1}=\overline{\widetilde{P}}(-x,y,t),\  \ \sigma \widetilde{L} \sigma^{-1}=\overline{\widetilde{L}}_{(x\rightarrow-x)}, \\
&&\sigma \widetilde{V}_{2}(x,y,t) \sigma^{-1}=\overline{\widetilde{V}}_{2}(-x,y,t),\  \ \sigma \widetilde{M} \sigma^{-1}=\overline{\widetilde{M}}_{(x\rightarrow-x)},
\end{eqnarray*}

which implies the solution for partially $\cal{PT}$-symmetric  nonlocal DS-I equation:
\begin{eqnarray}
 && \widetilde{u}=u+2 \alpha^{-1} S_{1,2}, \ \ \overline{\widetilde{u}}=\overline{u}+2 \epsilon  \alpha^{-1} S_{2,1}, \\
 && \widetilde{w}=w-2 \alpha^2 [\textbf{tr}(S)]_{x}=w-2 \alpha^2 [\ln(\det(\theta))]_{xx}.
\end{eqnarray}

In general, the $N$-fold Darboux matrix for partially $\cal{PT}$-symmetric nonlocal DS-I equation has the form:
\begin{eqnarray}\label{NDTDSI}
  T_{N}=\partial^{N}-\sum_{k=1}^{N}s_{k}\partial^{N-k}.
\end{eqnarray}

Transformation (\ref{NDTDSI}) maps: $L \rightarrow\tilde{L}=T_{N} L T_{N}^{-1}$,\  with\    $\tilde{L}=\partial_{y}-J\partial_{x}-P_{[N]}$, and the potential matrix has the relation:
\begin{eqnarray}
&& P_{[N]}=P+[J,s_{1}],  \label{DTPN} \\
&& V_{2,[N]}=V_{2}+V_{1,x}+2V_{0} s_{1,x}+[V_{0}, s_{1}]s_{1}+[V_{1}, s_{1}] \label{DTV2N} .
\end{eqnarray}

The coefficients matrices $s_{1}, s_{2}, \ldots ,s_{N}$ are determined by the system of linear algebraic equations:
\begin{equation}
T_{N}\left( \Psi_{k} \right)=0,\
\Psi_{k}=\left(
  \begin{array}{cc}
    \xi_{k}& -\epsilon \bar{\eta_{k}}  \\
    \eta_{k}& \bar{\xi_{k}} \\
  \end{array}
\right),\ k=1,2,...,N.
\end{equation}

Furthermore, the N-th order potential function for  partially $\cal{PT}$-symmetric  nonlocal DS-I equation solved from (\ref{DTPN})-(\ref{DTV2N}) can be represented in a determinant form:
\begin{eqnarray}\label{Wavefunctions}
 && u_{[N]}=u+2\alpha^{-1} \left(s_{1}\right)_{1,2},\  \  \overline{u}_{[N]}=\overline{u}+2\epsilon\alpha^{-1} \left(s_{1}\right)_{2,1},  \label{potentialu1}  \\
 && w_{[N]}=w-2\alpha^{-2} [\textbf{tr}(s_{1})]_{x}, \label{potentialw1}
\end{eqnarray}
where
\begin{eqnarray}
&&\left(s_{1}\right)_{1,2}=\det\Sigma^{1,2} \det\Sigma^{-1},\ \left(s_{1}\right)_{2,1}=\det\Sigma^{2,1} \det\Sigma^{-1},\
\Sigma=\left(
         \begin{array}{ccc}
           \partial^{N-1}\Psi_{1} & \cdots & \partial^{N-1}\Psi_{N} \\
          \cdots & \cdots & \cdots \\
            \Psi_{1} & \cdots & \Psi_{N} \\
         \end{array}
       \right),    \\
\nonumber &&\Sigma^{j,k} \ \text{is the matrix which derived by replacing the $k$-th row of} \ \Sigma \ \text{with the $j$-th row of}\
\left(
   \partial^{N}\Psi_{1}, \cdots, \partial^{N}\Psi_{N}
\right),\\
&&\left(s_{1}\right)_{k,j} \text{stands the  entry in the  $k$-th row and the $j$-th column of  matrix}  s_{1}.  \nonumber
\end{eqnarray}

Furthermore, (\ref{potentialw1}) can be further simplified  into another form:
\begin{eqnarray}
w_{[N]}=w-2 \alpha^{-2} [\ln(\det(\Sigma))]_{xx}, \label{potentialw}
\end{eqnarray}

and this can be verified via a direct calculation.

\subsection{Binary Darboux transformation for the partially $\cal{PT}$-symmetric nonlocal DS-II}
As we known, the local DS-I equation does not possess a Darboux transformation in differential form. Instead, it has a binary Darboux transformation in integral form.  As it has been shown for this  partially $\cal{PT}$-symmetric  nonlocal DS-I equation, we can construct an elementary DT in differential form, which has the same form with the DT reported in \cite{ZhouDS} where the DT is used to derive several types of bounded global explicit soliton solutions.    However,   for this  partially $\cal{PT}$-symmetric  nonlocal DS-II equation, the elementary DT is not enough. In the following, we are going to construct a  binary DT  in integral form for this equation.

Firstly, we recall some important properties for quasi-determinants which are introduced in Refs.\cite{GRdet1991,GelfandAdvace2005,Etingof1997,GNOYJPA2007, LCXNimmo2008}. It is a  generalization of  the determinant to matrices with noncommutative entries.  For a $n\times n$  matrix $M=(m_{i,j})$ over an, in general, non-commutative ring $\mathcal{R}$,     the  quasi-determinant for $M$ is defined by
\begin{eqnarray}
  |M|_{i,j} = m_{i,j}-r_{i}^{j}(M^{i,j})^{-1} c_{j}^{i} ,
\end{eqnarray}
where $r_{i}^{j}$ represents the $i$-th row of $M$ with the $j$-th element removed,  $c_{j}^{i}$ is the $j$-th column  of $M$ with the $i$-th element removed, and $M^{i,j}$ is a $(n-1)\times (n-1)$ minor obtained by deleting the  $i$-th row and the $j$-th column in $M$. Usually, as what is shown below, quasi-determinants can be  denoted  by boxing the entry about which the expansion is made
\begin{eqnarray}
 |M|_{i,j}=
\left|
\begin{array}{cc}
M^{i,j} & c_{j}^{i}  \\
 r_{i}^{j} & \fbox{$m_{i,j}$}
\end{array}
\right|.
\end{eqnarray}

In this paper, we consider the quasi-determinants that are only expanded about a term in the  last entry. Taking a block matrix
$M=\left(\begin{array}{cc}
A & B \\
C & d
\end{array}\right)$ for example, where $d \in \mathcal{R}$, $A$  is a  square matrix over $\mathcal{R}$ of arbitrary size, $B$, $C$ are column and row vectors over $\mathcal{R}$ with compatible lengths, respectively, then the quasi-determinant of $M$ expended about $d$ is
\begin{eqnarray*}
\left|
\begin{array}{cc}
A & B \\
 C & \fbox{$d$}
\end{array}
\right|= d-C A^{-1} B.
\end{eqnarray*}

Moreover,  as a quasi-determinant version of Jacobi¡¯s identity for determinants,  the noncommutative Sylvester¡¯s theorem was established in \cite{GRdet1991},
and a simple version of this theorem is given by
 \begin{eqnarray}\label{NCJacobi}
\left|
\begin{array}{ccc}
 E & F & G \\
 H & A & B \\
 J & C & \fbox{\emph{D}}
\end{array}
\right|=\left|
\begin{array}{cc}
 E & F  \\
 J & \fbox{\emph{D}}
\end{array}
\right|-\left|
\begin{array}{cc}
 E & F  \\
 J & \fbox{\emph{C}}
\end{array}
\right|\left|
\begin{array}{cc}
 E & F  \\
 J & \fbox{\emph{A}}
\end{array}
\right|^{-1}
\left|
\begin{array}{cc}
E & G  \\
H & \fbox{\emph{B}}
\end{array}
\right|.
 \end{eqnarray}

Next, we give a brief derivation of the DT for the partially $\cal{PT}$-symmetric nonlocal DS-II equation. For this equation, the operator $L$  has the constraint
\begin{eqnarray}\label{constraint}
 -\kappa  L^{\dag} \kappa^{-1}=L_{\left(x\rightarrow -x\right)},\ \kappa=\left(
                                                            \begin{array}{cc}
                                                              1 & 0 \\
                                                              0 & \epsilon \\
                                                            \end{array}
                                                          \right).
\end{eqnarray}
Here the denotation $L_{\left(x\rightarrow -x\right)}$ means changing all the variables $x$ in $L$ to $-x$. However, for this operator, one can not find a suitable matrix solution $\theta$  to construct the DT to preserve the constraint (\ref{constraint}). In order to overcome this problem one need to use the binary Darboux transformation (BDT). The standard BDT scheme has  been introduced and developed  in ref. \cite{MatveevDT1991}.  Several different forms of Darboux transformations  for the DS equations have been studied in Refs.\cite{LSYIP1992,GuManasPLA1996}. In this work, we inherit the  idea  from \cite{MatveevDT1991, NimmoGY2000}   and  construct a corresponding BDT for this partially $\cal{PT}$-symmetric nonlocal DS-II equation.

Considering operators $\hat{L}$, which is another copy of $L$ with new coefficients. Define the corresponding sets of non-singular solutions $\hat{S}$, let $\hat{\theta} \in \hat{S}$ s.t $G_{\hat{\theta}}: \hat{S}\rightarrow \tilde{S}$. Thus, we can get the following  mapping:
\begin{eqnarray*}
&& \ \ S \overset{G_{\theta}}{\longrightarrow} \tilde{S} \overset{G_{\hat{\theta}}^{-1}}{\longrightarrow} \hat{S}\\
&& S^{\dag} \overset{G_{\theta}^{\dag^{-1}}}{\longrightarrow} \tilde{S}^{\dag} \overset{G_{\hat{\theta}}^{\dag}}{\longrightarrow} \hat{S}^{\dag}
\end{eqnarray*}

For a given  $\phi \in S^{\dag}$, $G_{\theta}^{\dag^{-1}}(\phi) \in \tilde{S}^{\dag}$,  by determine the kernel of $G_{\hat{\theta}}^{\dag}$ we can obtain some nontrivial solutions in $\tilde{S}^{\dag}$. Thus, one can further define  a solution $\hat{\theta}=\left(G_{\theta}^{\dag^{-1}}(\phi)\right)^{\dag^{-1}}=-\theta \Omega^{-1}(\theta,\phi)$, and the BDT for operator $L$ is:
\begin{eqnarray}\label{BinaryDT}
G_{\theta, \phi} = G_{\hat{\theta}}^{-1} G_{\theta}=I -\theta \Omega^{-1}(\theta,\phi)\partial^{-1}\phi^{\dag}, \ \ \ \Omega(\theta,\phi)=\partial^{-1} ( \phi^{\dag} \theta ).
\end{eqnarray}

To proceed the iteration of DT we also need
\begin{eqnarray}\label{InverseBDT}
G_{\theta, \phi}^{\dag^{-1}} = G_{\hat{\theta}}^{\dag} G_{\theta}^{\dag^{-1}}=I - \phi \Omega^{\dag^{-1}}(\theta,\phi)\partial^{-1}\theta^{\dag}.
\end{eqnarray}

This Darboux transformation makes sense for any $m\times k$ matrices $\theta$ and $\phi$, and we only need   $\Omega(\theta,\phi)$ to be an invertible square matrix.
To reduce (\ref{BinaryDT})  to the BDT for the  partially $\cal{PT}$-symmetric nonlocal DS-II equation,  we have to take the choice according to symmetry (\ref{constraint}) that: $\phi(x,y,t)=R^{\dag}(\theta\left(-x,y,t\right)),\ R=-i\kappa$. Then the potential solutions in this equation can be constructed by the  combination of an elementary DT with its inverse:
\begin{eqnarray}
 && \hat{P}=P+ [J, \theta \Omega^{-1}(\theta,\phi)\phi^{\dag}], \\
 && \hat{w}=w - 2 [\textbf{tr}(\theta \Omega^{-1}(\theta,\phi)\phi^{\dag})]_{x}.
\end{eqnarray}

The above Binary DT is iterated as following:
\begin{eqnarray}
  &&\Phi_{[n+1]}=G_{\theta_{[n]}, \phi_{[n]}}\left(\Phi_{[n]}\right)=\Phi_{[n]}-\theta_{[n]}\Omega^{-1}(\theta_{[n]},\phi_[n])\Omega(\Phi_{[n]},\phi_[n]), \\
  &&\Psi_{[n+1]}=G_{\theta_{[n]}, \phi_{[n]}}^{\dag^{-1}}\left(\Psi_{[n]}\right)=\Psi_{[n]}-\phi_{[n]}\Omega^{\dag^{-1}}(\theta_{[n]},\phi_[n])\Omega(\Psi_{[n]},\theta_{[n]}),  \\
  && \theta_{[n]}=\lim_{\Phi \rightarrow \theta_{n}}\Phi_{[n]},\ \  \phi_{[n]}=\lim_{\Psi \rightarrow \phi_{n}} \Psi_{[n]}.
\end{eqnarray}

For the potential matrix,  introducing a $2 \times 2$ matrix $Q$ s.t $P=[Q, J]$, which of the form:
\begin{eqnarray}
 Q=-\frac{\alpha}{2} \left(
     \begin{array}{cc}
       * & u(x,y,t) \\
       \epsilon \bar{u}(-x,y,t) & * \\
     \end{array}
   \right),
\end{eqnarray}
while the entries * are arbitrary and do not contribute to $P$. Then it follows from (22) that:
\begin{eqnarray}
  &&\hat{P}=[\hat{Q}, J], \\
  &&\hat{Q}=Q-\theta \Omega^{-1}(\theta,\phi)\phi^{\dag}.
\end{eqnarray}

After n times applications of the BDT we obtain:
\begin{eqnarray}
 && Q_{[n+1]}=Q_{[n]}-\theta_{[n]} \Omega^{-1}(\theta_{[n]},\phi_{[n]})\phi_{[n]}^{\dag},  \label{BDTQS} \\
 && w_{[n+1]}=w_{[n]} - 2 [\textbf{tr}(\theta_{[n]} \Omega^{-1}(\theta_{[n]},\phi_{[n]})\phi_{[n]}^{\dag})]_{x}. \label{BDTWS}
\end{eqnarray}

Denoting
\begin{eqnarray*}
\Theta=\left( \theta_{1},\cdots, \theta_{n} \right),\ \textbf{P}=\left( \phi_{1},\cdots, \phi_{n} \right),\
W=\left(
    \begin{array}{cc}
      \frac{\partial^{-1}(w)}{2} & 0 \\
      0 & 0 \\
    \end{array}
  \right).
\end{eqnarray*}

By using the noncommutative Jacobi identity (\ref{NCJacobi}), one can express the  above results on $n$-th order BDT in terms of quasi-determinants:
\begin{eqnarray}
&&\Phi_{[n+1]}=\left|
\begin{array}{cc}
 \Omega(\Theta,\textbf{P}) & \Omega(\Phi_{[1]},\textbf{P}) \\
 \Theta  & \fbox{$\Phi_{[1]}$}
  \\
\end{array}
\right|,\ \
\Psi_{[n+1]}=\left|
\begin{array}{cc}
 \Omega^{\dag}(\Theta,\textbf{P}) & \Omega^{\dag}(\Theta,\Psi_{[1]}) \\
 \textbf{P}& \fbox{$\Psi_{[1]}$}
  \\
\end{array}
\right|,\\
&&Q_{[n+1]}=\left|
\begin{array}{cc}
 \Omega(\Theta,\textbf{P}) & \textbf{P}^{\dag} \\
 \Theta & \fbox{$Q_{[1]}$}
  \\
\end{array}
\right|,\
w_{[n+1]}=2\partial_{x}[\textbf{tr}\left(\left|
\begin{array}{cc}
 \Omega(\Theta,\textbf{P}) & \textbf{P}^{\dag} \\
 \Theta & \fbox{$W$}
  \\
\end{array}\right|
\right)]. \label{Nsolitons}
\end{eqnarray}

For convenience, introducing vectors $\psi_{i}$ and $\varphi_{i}$, $i=1,2$, which satisfy:
\begin{eqnarray*}
   \Theta=\left(
            \begin{array}{c}
              \psi_{1} \\
               \psi_{2} \\
            \end{array}
          \right),\ \ \
           \textbf{P}=\left(
            \begin{array}{c}
              \varphi_{1} \\
               \varphi_{2} \\
            \end{array}
          \right),
\end{eqnarray*}

we further obtain :
\begin{eqnarray}\label{quasi-Q}
Q_{[n+1]}=Q_{[1]}+\left(\begin{array}{cc}
\left|
\begin{array}{cc}
 \Omega(\Theta,\textbf{P}) & \varphi_{1}^{\dag} \\
 \psi_{1} & \fbox{0}
\end{array}
\right| &
\left|
\begin{array}{cc}
 \Omega(\Theta,\textbf{P}) & \varphi_{2}^{\dag} \\
 \psi_{1} & \fbox{0}
\end{array}
\right| \\
\left|
\begin{array}{cc}
 \Omega(\Theta,\textbf{P}) & \varphi_{1}^{\dag} \\
 \psi_{2} & \fbox{0}
\end{array}
\right| &
\left|
\begin{array}{cc}
 \Omega(\Theta,\textbf{P}) & \varphi_{2}^{\dag} \\
 \psi_{2} & \fbox{0}
\end{array}
\right|
 \end{array}
\right).
\end{eqnarray}

Then, combination of (\ref{BDTQS})-(\ref{BDTWS}) and (\ref{quasi-Q}) leads to the transformations  between  potential functions in terms of quasi-grammian expressions:
\begin{eqnarray}
&&u_{[n+1]}(x,y,t)= u(x,y,t)-\frac{2}{\alpha} \left|
\begin{array}{cc}
 \Omega(\Theta,\textbf{P}) & \varphi_{2}^{\dag} \\
 \psi_{1} & \fbox{0}
\end{array}
\right|,  \label{un+1} \\
&&w_{[n+1]}(x,y,t)= w(x,y,t)+2\partial_{x}[\textbf{tr}\left(\left|
\begin{array}{cc}
 \Omega(\Theta,\textbf{P}) & \textbf{P}^{\dag} \\
 \Theta & \fbox{$\mathbf{0}$}
  \\
\end{array}\right|
\right)].\label{wn+1}
\end{eqnarray}

Here $\mathbf{0}$  represents  a  $2\times2$ zero matrix. Noting that in this expression one has to calculate the inverse of matrix $\Omega(\Theta,\textbf{P})$. To overcome this problem, we can reformulate the expression  into the quotient of determinants instead of using quasi determinants, that is
\begin{eqnarray}\label{quasidet2det}
\left|
\begin{array}{cc}
 \Omega(\Theta,\textbf{P}) & \varphi_{j}^{\dag} \\
 \psi_{k} & \fbox{0}
\end{array}
\right|=\frac{\det\left(M^{(k,j)}\right)}{\det(\Omega(\Theta,\textbf{P}))}, \ \text{where}\
M^{(k,j)}=\left(
\begin{array}{cc}
 \Omega(\Theta,\textbf{P}) & \varphi_{j}^{\dag} \\
 \psi_{k} & 0
\end{array}
 \right), 1\leq k, j \leq2.
\end{eqnarray}

\textbf{Remark 2.2.1} \emph{Moreover,  formula (\ref{wn+1})  can be further transformed  into a more compact form:}
\begin{eqnarray}
w_{[n+1]}(x,y,t)= w(x,y,t) - 2\partial_{x}^{2}\{\log \left[\det\left( \Omega(\Theta,\textbf{P})\right) \right] \}.
\end{eqnarray}

\emph{Proof.} \   In fact, via the Laplace expansion into the \textbf{trace} in (35), we can show that
\begin{eqnarray*}
 \textbf{tr}\left(\left|
\begin{array}{cc}
 \Omega(\Theta,\textbf{P}) & \textbf{P}^{\dag} \\
 \Theta & \fbox{$\mathbf{0}$}
  \\
\end{array}\right|
\right)
= \frac{\sum _{k=1}^n \sum _{j=1}^n (-1)^{j+k-1}  \left(\phi _j^{\dag} \theta _k\right)M_{j,k}}{\det\left( \Omega(\Theta,\textbf{P})\right)},
\end{eqnarray*}

where $M_{j,k}$ is the minor matrix of $ \Omega(\Theta,\textbf{P}) $, i.e., the determinant of a $(n-1)\times (n-1)$ matrix that results

from the $j$-th row and the $k$-th column of $\left( \Omega(\Theta,\textbf{P})\right)$.
On the other hand, it  can be  easily verified  that
\begin{eqnarray*}
\partial_{x}\{\det\left( \Omega(\Theta,\textbf{P})\right)\}=
\sum _{k=1}^n \sum _{j=1}^n (-1)^{j+k}  \left(\phi _j^{\dag} \theta _k\right)M_{j,k}.
\end{eqnarray*}

And this completes the proof.

\subsection{High-order Darboux transformation for the  partially $\cal{PT}$-symmetric  nonlocal DS system}
To construct the high-order solution, the high-order Darboux transformation are needed. It is assumed by introducing a parameter $k_{i}$ in the fundamental  matrix solution $\theta_{i}(k_{i})$. As it was pointed in Ref\cite{MatveevDT1991}, a generalized DT does exist. Through a limiting process, the general high-order DT for nonlocal DS-I equation is constructed in the following forms:

\emph{\textbf{Theorem 1}} (Theorem 2, \cite{LLMPRE2012})\ \ \ \emph{Assuming $ \Psi_{i}(k_{i}),\ i=1,2,...,n$ (which are given in (\ref{Wavefunctions}))  are $n$ distinct matrix solutions of the linear problem (\ref{Lax-pair1})-(\ref{Lax-pair2}), and their Taylor expansions are}
\begin{eqnarray*}
&&\Psi_{i}(k_{i}+\delta)=\Psi_{i}(k_{i})+\Psi_{i}^{[1]}\delta+\cdots+\Psi_{i}^{[m_{i}]}\delta^{m_{i}}+\cdots,\ i=1,2,\ldots,n,\\
&&\Psi_{i}^{[j]}=\frac{1}{j!}\frac{\partial^{j}}{\partial k^{j}}\Psi_{i}(k)|_{k=k_{i}},\ j=1,2,\cdots.
\end{eqnarray*}
\emph{Then the $N$-fold generalized Darboux transformation is defined as}
\begin{eqnarray*}
T=G_{n}G_{n-1}\cdots G_{0},
\end{eqnarray*}
\emph{where,}
\begin{align*}
G_{i}=&G_{i}[m_{i}]\cdots G_{i}[1]\ (i\geq1),\ G_{0}=I,\ n+ \sum_{i=1}^{n}m_{i}=N,\\
G_{i}[j]&=\partial_{x}- \Psi_{i,x}[j-1] \Psi_{i}[j-1]^{-1},\ \ 1 \leq j \leq m_{i},\\
\Psi_{i}[k]&=\lim_{\delta\rightarrow0}\frac{[G_{i}[k]\cdots G_{i}[1] G_{i-1}\cdots G_{0}]}{\delta^{k}} \Psi_{i}(k_{i}+\delta),\\
&=G_{i}[k]\cdots G_{i}[1] G_{i-1}\cdots G_{0}[ \Psi_{i}^{[k]}(k_{i})].
\end{align*}
\emph{By performing the above limit process on the determinant form (\ref{potentialu1}), we get the formula for high-order solutions for the  partially $\cal{PT}$-symmetric  nonlocal DS-I equation:
}\begin{eqnarray}
&& u_{[N]}=u+2\alpha^{-1} \det\Sigma_{0}^{1,2} \left(\det\Sigma_{0}\right)^{-1},\label{GDTpotentialu}\\
&& w_{[N]}=w-2 \alpha^{-2} [\ln(\det(\Sigma_{0}))]_{xx}, \label{GDTpotentialw}
\end{eqnarray}
\emph{where,}
\begin{eqnarray*}
 \Sigma_{0}=[\Sigma_{1}\ ...\ \Sigma_{n}],\ \ \Sigma_{0}^{1,2}=[\Sigma_{1}^{1,2}\ ...\ \Sigma_{n}^{1,2}],\
  \Sigma_{i}=\left(
         \begin{array}{ccc}
           \partial^{N-1}\Psi_{i} & \cdots & \partial^{N-1}\Psi_{i}^{[m_{i}]} \\
          \cdots & \cdots & \cdots \\
            \Psi_{i} & \cdots & \Psi_{i}^{[m_{i}]} \\
         \end{array}
       \right),
\end{eqnarray*}
and\ $\Sigma_{i}^{j,k} \ \text{\emph{is the matrix which derived by replacing the $k$-th row of}} \ \Sigma_{i} \ \text{\emph{with the $j$-th row of}}\
\left(
   \partial^{N}\Psi_{i}, \cdots, \partial^{N}\Psi_{i}^{[m_{i}]}
\right)$.

Next,  following  the idea proposed for the nonlinear Schr$\ddot{o}$dinger equation in \cite{LLMPRE2012}, we construct the corresponding high-order DT in the partially $\cal{PT}$-symmetric  nonlocal DS-II equation.  Indeed, the binary DT considered above are degenerate  in the sense that $G_{\theta_{1}, \phi_{1}}(\theta_{1})=0$ and  $G_{\theta_{1}, \phi_{1}}^{\dag^{-1}}(\phi_{1})=0$, thus we may work with
\begin{eqnarray*}
\theta_{1}[1]=\lim_{\delta \rightarrow 0}\frac{G_{\theta_{1}, \phi_{1}}(\theta_{1}(k_{1}+\delta))}{\delta}=G_{\theta_{1}, \phi_{1}}\frac{d\theta_{1}}{dk}|_{k=k_{1}},\ \
\phi_{1}[1]=\lim_{\tilde{\delta} \rightarrow 0}\frac{G_{\theta_{1}, \phi_{1}}^{\dag^{-1}}(\phi_{1}(\tilde{k}_{1}+\tilde{\delta}))}{\tilde{\delta}}=G_{\theta_{1}, \phi_{1}}^{\dag^{-1}}\frac{d\phi_{1}}{d\tilde{k}}|_{\tilde{k}=\tilde{k}_{1}}.
\end{eqnarray*}

This serves the seed solution for proceeding the next step binary Darboux transformation. Generally, we
assume that solutions $\theta_{i}=\left(\xi_{i},\ \eta_{i}\right)^{T}\ (i=1\ldots s)$
are given for the Lax operator $L$ and  solutions $\rho_{i}=\left(\mu_{i},\ \nu_{i}\right)^{T}\ (i=1\ldots s) $ are given for its adjoint operator $L^{\dag}$, then we have the following generalized Binary DT.

\emph{\textbf{Theorem 2}} \ \ \emph{Let solutions $\left(\xi_{i},\ \eta_{i}\right)^{T}\in S$, and $\left(\mu_{i},\ \nu_{i}\right)^{T}\in S ^{\dag}\ (i=1\ldots s)$, so the high-order Binary DT is constructed in the form  as}
\begin{eqnarray*}
  G_{N}=G_{\theta_{s}^{[m_{s}-1]},\ \rho_{s}^{[m_{s}-1]}} \cdots  G_{\theta_{s},\ \rho_{s}} \cdots
  G_{\theta_{1}^{[m_{1}-1]},\ \rho_{1}^{[m_{1}-1]}}\cdots G_{\theta_{1},\ \rho_{1}},
\end{eqnarray*}
\emph{where  $N=\sum_{i=1}^{s}m_{i}$, and}
\begin{eqnarray*}
G_{\theta_{i}^{[j]},\ \rho_{i}^{[j]}}=I-\theta_{i}^{[j]}\Omega^{-1}\left(\theta_{i}^{[j]},\ \rho_{i}^{[j]}\right)\partial^{-1} \rho_{i}^{[j]\dag},\\
G^{\dag^{-1}}_{\theta_{i}^{[j]},\ \rho_{i}^{[j]}}=I-\rho_{i}^{[j]}\Omega^{-1}\left(\rho_{i}^{[j]},\ \theta_{i}^{[j]}\right)\partial^{-1} \theta_{i}^{[j]\dag},
\end{eqnarray*}
\emph{here $\theta_{i}^{[j]}$  and $\rho_{i}^{[j]}$ are derived by performing the limit on the fundamental eigenfunctions with perturbation parameters  $\delta$ and $\tilde{\delta}$:}
\begin{eqnarray*}
\theta_{i}^{[j]}=\lim_{\delta\rightarrow0}\frac{\left[G_{\theta_{i}^{[j-1]},\ \rho_{i}^{[j-1]}} \cdots  G_{\theta_{i},\ \rho_{i}} \cdots
  G_{\theta_{1}^{[m_{1}-1]},\ \rho_{1}^{[m_{1}-1]}}\cdots G_{\theta_{1},\ \rho_{1}}\right]_{k=k_{i}+\delta}\theta_{i}(k_{i}+\delta)}{\delta^{j}},\\
\rho_{i}^{[j]}=\lim_{\tilde{\delta}\rightarrow0}\frac{\left[G^{\dag^{-1}}_{\theta_{i}^{[j-1]},\ \rho_{i}^{[j-1]}} \cdots  G^{\dag^{-1}}_{\theta_{i},\ \rho_{i}} \cdots
  G^{\dag^{-1}}_{\theta_{1}^{[m_{1}-1]},\ \rho_{1}^{[m_{1}-1]}}\cdots G^{\dag^{-1}}_{\theta_{1},\ \rho_{1}}\right]_{k=\tilde{k_{i}}+\tilde{\delta}}\rho_{i}(\tilde{k_{i}}+\tilde{\delta})}{\tilde{\delta}^{j}}.
\end{eqnarray*}

By taking above limitation  directly on (\ref{un+1})-(\ref{wn+1}), the transformations between potential matrices can be represented in a form of quasi-gram determinant.

\emph{\textbf{Theorem 3}} \ \ \
\emph{The above generalized binary Darboux  matrix and the corresponding transformation between the potential matrices can be represented as the following forms:
}\begin{eqnarray}
&&Q_{[N]}=Q_{[1]}+\left|
\begin{array}{cc}
 \Omega(\Theta,\textbf{P}) & \textbf{P}^{\dag} \\
 \Theta & \fbox{$0$}
  \\
\end{array}
\right|=Q_{[1]}- \Theta  \Omega^{-1}(\Theta,\textbf{P})\textbf{P}^{\dag}, \label{gDTDS-IIQN}\\
&& w_{[N]}(x,y,t)= w_{[1]}(x,y,t) - 2\partial_{x}^{2}\{\log \left[\det\left( \Omega(\Theta,\textbf{P})\right) \right] \},\label{gDTDS-IIwN}
\end{eqnarray}
\emph{where,}
\begin{eqnarray*}
&&\theta_{i}=\theta_{i}(k_{i}+\delta),\ \rho_{j}=\rho_{j}(\tilde{k_{i}}+\tilde{\delta}),\\
 && \Theta=\left( \Theta_{1}, \Theta_{2},\ldots, \Theta_{s} \right),\ \Theta_{i}=\left(\theta_{i}, \frac{d \theta_{i}}{d \delta},\ldots, \frac{1}{(r_{i}-1)!}\frac{d^{ r_{i}-1} \theta_{i}}{d \delta^{r_{i}-1}} \right)|_{\delta \rightarrow 0},\\
 && \textbf{P}=\left( \textbf{P}_{1}, \textbf{P}_{2},\ldots, \textbf{P}_{s} \right),\ \textbf{P}_{j}=\left(\rho_{j}, \frac{d \rho_{j}}{d \tilde{\delta}} ,\ldots, \frac{1}{(r_{j}-1)!} \frac{d^{r_{j}-1}  \rho_{j}}{d \tilde{\delta}^{r_{j}-1}} \right)|_{\tilde{\delta}\rightarrow 0},\\
 && \Omega(\Theta,\textbf{P})=\left( \Omega^{[i j]}\right)_{1\leq i,j \leq s}, \ \Omega^{[ij]}=\left( \Omega^{[ij]_{m,n}} \right)_{r_{i}\times r_{j}}, \\
&& \Omega^{[ij]_{m,n}}=\lim_{\delta,\  \tilde{\delta} \rightarrow 0}\frac{1}{(m-1)!(n-1)!}\frac{\partial^{m+n-2}}{\partial\delta^{n-1}\partial\tilde{\delta}^{m-1}}\Omega(\theta_{j}, \rho_{i}).
\end{eqnarray*}

\emph{Proof}: The above results can be obtained  by directly taking limits in formula (\ref{Nsolitons}) with property (\ref{quasidet2det}).

 One can further derive the Buckl\"{u}nd transformation of solution $u_{[N]}(x,y,t)$ from (\ref{gDTDS-IIQN}), which is taken from the 1-st row and the 2-nd column element in potential matrix $Q_{[N]}$.

\section{General rational solution in partially $\cal{PT}$-symmetric  nonlocal DS-I system}

It is shown in ref.\cite{YangDSI,YangDSII} that with the bilinear method,  a family of  rational solutions lead to the rogue waves for the local DS equations. In this work, the rogue wave solution for nonlocal DS equations was derived via a generalized version of Darboux transformation.

The general form of eigenfunctions are solved from the system (\ref{Lax-pair1})-(\ref{Lax-pair2}) when  the initial potential solution $u$ is taken  as a real constant $\rho$, which of the form:
\begin{eqnarray*}
 &&\xi_{i}(x,y,t)=\rho_i \exp \left[\omega _i(x,y,t)\right],\\
  &&\eta_{i}(x,y,t)=\frac{ \lambda_i \rho_i }{\rho }\exp \left[\omega _i(x,y,t)\right],\\
  && \omega _i(x,y,t) =\alpha_{i} x+ \beta_{i} y+\gamma_{i} t,\\
&& \alpha_{i}=-\frac{1}{2} \alpha  \left(\lambda_{i} +\frac{\epsilon \rho^2 }{\lambda_{i} }\right),\   \beta_{i}=\frac{1}{2} \left(\lambda_{i} -\frac{\epsilon \rho^2 }{\lambda_{i} }\right),\ \gamma_{i}= \textrm{i} \alpha^{-1}  \alpha_{i}  \beta_{i},
\end{eqnarray*}
where  $\lambda_{i}=r_i \exp \left(\textrm{i} \varphi _i \right)$, $r_i$, $\varphi _i$ and $\rho$ are free real parameters, $\rho_{i}$ is set to be  complex.

Generally, to derive rational type solutions, we choose the eigenfunction via superposition principle, which can be written   in the form as:
\begin{eqnarray}\label{Eigenfunctions}
\{\mathcal{F}_k +\partial_{ \varphi_{k}}\}
\left(
 \xi _k,
 \eta _k
\right)^T :=
\left(
            \begin{array}{c}
            \mathit{P}_{k}(x,y,t)\xi _k \\
             \mathit{Q}_{k}(x,y,t) \eta _k \\
            \end{array}
          \right),
\
\mathcal{F}_k=\mathit{e_{k}}+\textrm{i} \mathit{f_{k}},\ \left(e_{k}, f_{k}\in \mathbb{R}\right),
\end{eqnarray}
where,
\begin{eqnarray*}
&&\mathit{P}_{k}(x,y,t)=\mathcal{F}_k+\rho_k^{-1}\rho_{k,\varphi_{k}}+\left(-\textrm{i} \alpha\beta_{k}\right) x+ \left(-\textrm{i} \alpha^{-1} \alpha_{k} \right) y+\frac{1}{2} \left( \lambda_{k}^2+\frac{\rho^4}{\lambda_{k}^2} \right) t;\\
&&\mathit{Q}_{k}(x,y,t)=\mathcal{F}_k+ \lambda_k   \rho^{-1} \left(\textrm{i}\rho_{k}+\rho_{k,\varphi_{k}} \right)+\left(-\textrm{i} \alpha\beta_{k}\right) x+ \left(-\textrm{i} \alpha^{-1} \alpha_{k} \right) y+\frac{1}{2} \left( \lambda_{k}^2+\frac{\rho^4}{\lambda_{k}^2} \right) t;
\end{eqnarray*}

\subsection{Fundamental  rogue-wave in partially $\cal{PT}$-symmetric  nonlocal DS-I}
 To derive the first order rational solution,  we set $N=1$, $\rho=1$ with $\rho_{1}=\exp \left(\frac{-\textrm{i} \varphi _1}{2}\right) $ in formula (\ref{potentialu1})-(\ref{potentialw1}). Then the  first-order rational solution is
\begin{eqnarray}
&&u_{1}(x,y,t)=1-\frac{2i \emph{F}_{1}(x,y,t)+1}{\emph{F}(x,y,t)},  \label{1-stRwu}\\
&&w_{1}(x,y,t)=\epsilon-2  [\ln(F(x,y,t))]_{xx}, \label{1-stRww}
\end{eqnarray}
where,
\begin{eqnarray}
 \nonumber
&&\emph{F}(x,y,t)= \emph{F}_{1}^2(x,y,t)+\emph{F}_{2}^2(x,y,t)+\frac{\epsilon r_{1}^2}{( \epsilon +r_{1}^2)^2}, \
\emph{p}_{1}=\frac{ r_{1}-\epsilon r_{1}^{-1}}{2},\ \emph{q}_{1}=\frac{ r_{1}+\epsilon r_{1}^{-1}}{2},
\\  \nonumber
&&\emph{F}_{1}(x,y,t)=-i\emph{p}_{1} x \cos \varphi_{1} -  \emph{p}_{1} y \sin \varphi_{1} +  (\emph{p}_{1}^{2}+  \emph{q}_{1}^{2}) t \cos 2\varphi_{1} + e_{1}, \\
&&\nonumber \emph{F}_{2}(x,y,t)=i\emph{q}_{1}x \sin \varphi_{1} - \emph{q}_{1} y \cos \varphi_{1}  - 2 \emph{p}_{1} \emph{q}_{1} t \sin 2\varphi_{1}+ \frac{\emph{p}_{1}}{2\emph{q}_{1}}+\mathit{f_{1}}.
\end{eqnarray}
By analysing the denominator in solution (\ref{1-stRwu}), it is shown that this  rational solution has different dynamical  patterns according to the parameter values of $ r_{1}$ and $\varphi_{1}$.

(i). If $\varphi_{1}=k\pi$ ($k=0, \pm1, \pm2,...$), then  $\lambda_{1}=(-1)^n r_{1}$ is a real number. In this case, it is a rogue wave which approaches a constant background, i.e., $u_{1}\rightarrow 1$, $w_{1}\rightarrow \epsilon$ as $t\rightarrow -\infty$. And the function $F$ in solution (\ref{1-stRwu}) becomes
\begin{eqnarray*}
F(x,y,t)=\left[\textrm{i}\emph{p}_{1} (-1)^{n+1} x  + (\emph{p}_{1}^{2}+  \emph{q}_{1}^{2}) t  + e_{1}\right]^2+\left[ (-1)^{n+1}\emph{q}_{1} y + \frac{\emph{p}_{1}}{2\emph{q}_{1}}+\mathit{f_{1}}\right]^2+\frac{\epsilon r_{1}^2}{( \epsilon +r_{1}^2)^2}.
\end{eqnarray*}

This function becomes zero  at a critical time $ t_{c}=\frac{-2\mathit{e_{1}} r_{1}^2}{ 1+r_{1}^4 }$, and it
occurs  on the $(x,y)$ plane when $r_{1}^2\neq1$:
\begin{eqnarray*}
-\emph{p}_{1}^2 x^2+\left[ (-1)^{n+1}\emph{q}_{1} y + \frac{\emph{p}_{1}}{2\emph{q}_{1}}+\mathit{f_{1}}\right]^2+\frac{\epsilon r_{1}^2}{( \epsilon +r_{1}^2)^2}=0.
\end{eqnarray*}
Thus, this rogue wave arises from a constant background and develops finite-time singularity on a hyperbola at $t_{c}=\frac{-2\mathit{e_{1}} r_{1}^2}{ 1+r_{1}^4 }$,
and it shows some cross-shape properties at some time points. For example, if we take $\varphi_{1}=2\pi$, $r_{1}=2$ with $f_1=1,\ e_1=0$, the  singularity of this solution occurs when $t=0$.  Here we only plot solutions up to time $t =-0.03$ in  Fig.1, shortly before the exploding time, where the amplitude of rogue wave could attain very high.
\begin{figure}[htb]
   \begin{center}
   \includegraphics[scale=0.280, bb=0 0 385 567]{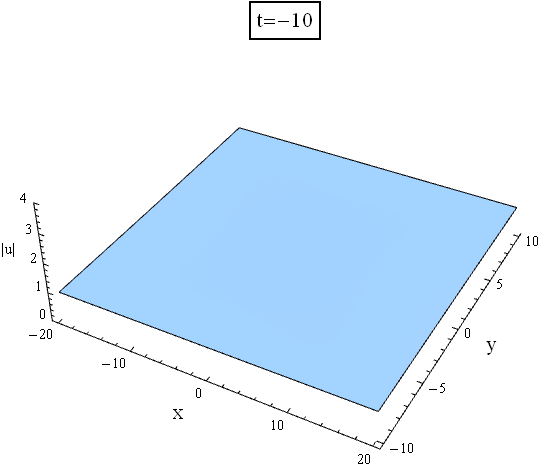}\hspace{2.5cm}
   \includegraphics[scale=0.280, bb=0 0 385 567]{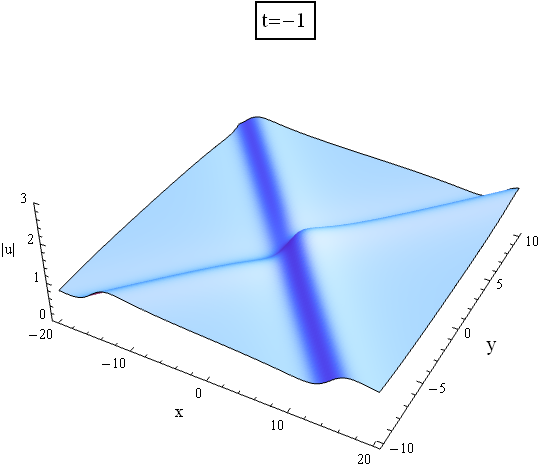} \\
   \includegraphics[scale=0.280, bb=0 0 385 567]{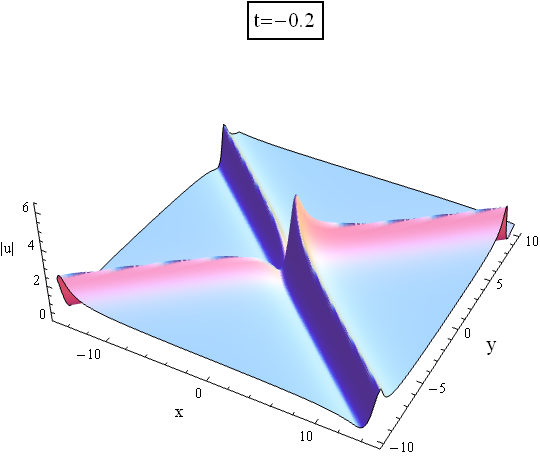}\hspace{2.5cm}
      \includegraphics[scale=0.280, bb=0 0 385 567]{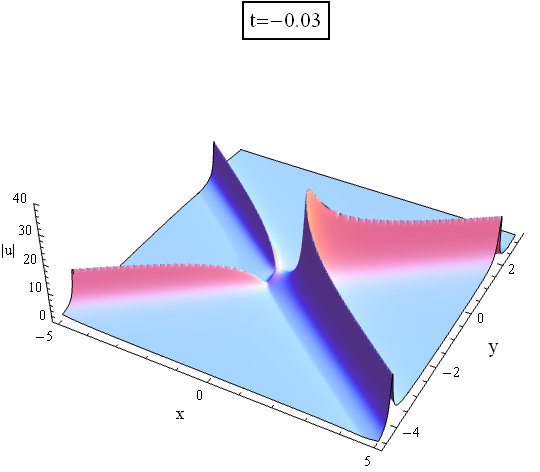}
   \caption{Time evolution for the 1-st order cross-shape exploding fundamental rogue wave solution in the nonlocal DS-I equation,
       with parameters: $\varphi_{1}=2\pi$, $r_{1}=2$, $e_1=0$, $f_1=1$.}
   \label{picture-label}
   \end{center}
\end{figure}

Moreover, in the de-focusing case $\epsilon=-1$, for any $t_{c}\in I_{c}$,  where
\begin{eqnarray*}
 I_{c}=\left[ \left(-\frac{|r_{1}|}{|r_{1}^2-1|}-e_{1}\right)\frac{2}{(r_{1}^2-r_{1}^{-2})},\ \left(\frac{|r_{1}|}{|r_{1}^2-1|}-e_{1}\right)\frac{2}{(r_{1}^2-r_{1}^{-2})}\right],\ r_{1}^2\neq1,
\end{eqnarray*}
function $F(x,y,t)$ becomes zero at the spatial locations $x=0, y=y_{\pm c}$, and $y_{\pm c}$  are solved from the following  quadratic equation:
\begin{eqnarray*}
\left[ (-1)^{n+1}\emph{q}_{1} y + \frac{\emph{p}_{1}}{2\emph{q}_{1}}+\mathit{f_{1}}\right]^2=\frac{ r_{1}^2}{( r_{1}^2-1 )^2}-\left[ \left(\frac{r_{1}^2+r_{1}^{-2}}{2}\right) t_{c}  + e_{1}\right]^2.
\end{eqnarray*}
Therefore, when $\epsilon=-1$, this rogue wave develops extra singularity  on a finite-time interval.

In addition, as a special case, when $r_{1}=1$, $\epsilon=1$, this rogue wave is $x$-independent and degenerates into the following Peregrine soliton for the nonlocal NLS equation
\begin{eqnarray}
 u_{1}(x,y,t)=1-\frac{2 i t+2 i \mathit{e}_{1}+1}{\left(y \pm \mathit{f}_{1}\right)^2+ \left( t+\mathit{e}_{1} \right)^2+\frac{1}{4}},
\end{eqnarray}
 where the parameters $\mathit{e}_{1}$ and $\mathit{f}_{1}$ can be moved by a shifting. Besides, this solution, in terms of the (1+2) dimensional space, is a (1+1) dimensional   line rogue wave in this nonlocal  DS-I equation, see Fig.2.

As a trivial case, when $r_{1}=1$  and $\epsilon=-1$, then  $ u_{1}(x,y,t) \rightarrow 1$.
\begin{figure}[htbp]
  \centering
   \includegraphics[scale=0.255, bb=0 0 385 567]{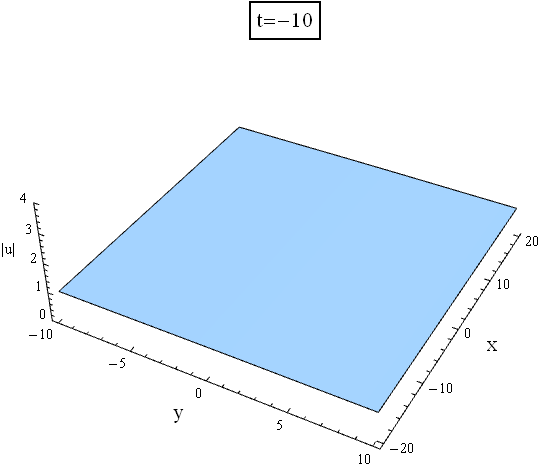}\hspace{2.0cm}
    \includegraphics[scale=0.255, bb=0 0 385 567]{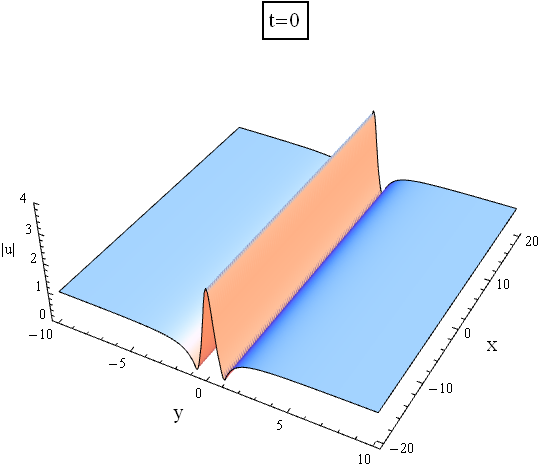}\hspace{2.0cm}
   \includegraphics[scale=0.255, bb=0 0 385 567]{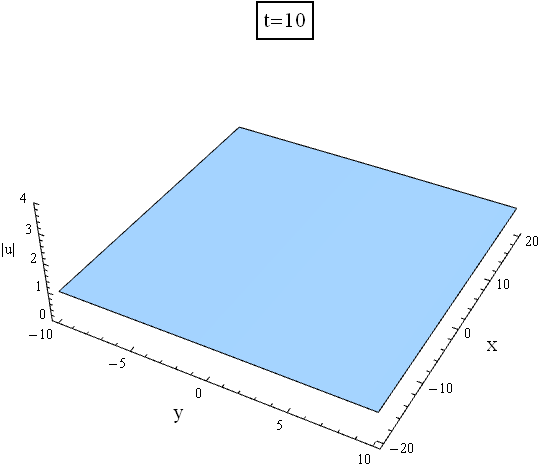}
   \caption{Fundamental (1+1)-dimensional line rogue wave  in the nonlocal DS-I equation, with parameters: $f_1=0,\ e_1=0$.}
   \label{picture-label}
\end{figure}

(ii). In another case, when $\epsilon=1$, $\varphi_{1}=\frac{(2k-1)\pi}{2}$, i.e., $\lambda$ is a purely imaginary. It can generate a two-dimensional non-singular  rational travelling wave solution(while $\epsilon=-1$ may cause some singularities).  The ¡°ridge¡± of the solution lays approximately on the following $[x(t), y(t)]$ trajectory:
\begin{center}
  $(-1)^{k-1}\frac{\left(1+r_{1}^2\right)}{r_{1}} x + (-1)^{k-1} \frac{\left(1-r_{1}^2 \right)}{r_{1}} y-\frac{1+r_{1}^4}{r_{1}^2}t+e_{1}=0$,\\ $(-1)^{k}\frac{\left(1+r_{1}^2\right)}{r_{1}} x + (-1)^{k-1} \frac{\left(1-r_{1}^2 \right)}{r_{1}} y-\frac{1+r_{1}^4}{r_{1}^2}t+e_{1}=0$.
\end{center}
\begin{figure}[htb]
   \begin{center}
   \includegraphics[scale=0.255, bb=0 0 385 567]{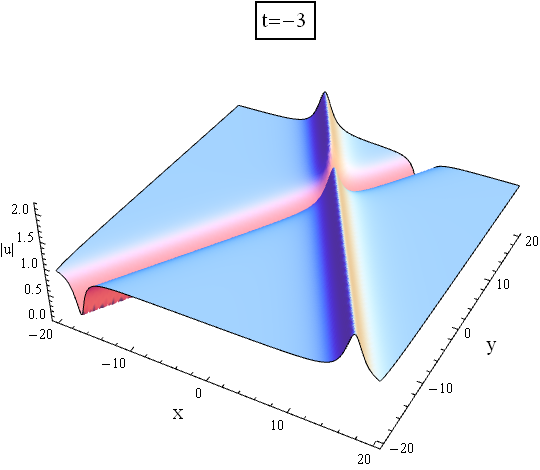}\hspace{2.0cm}
   \includegraphics[scale=0.255, bb=0 0 385 567]{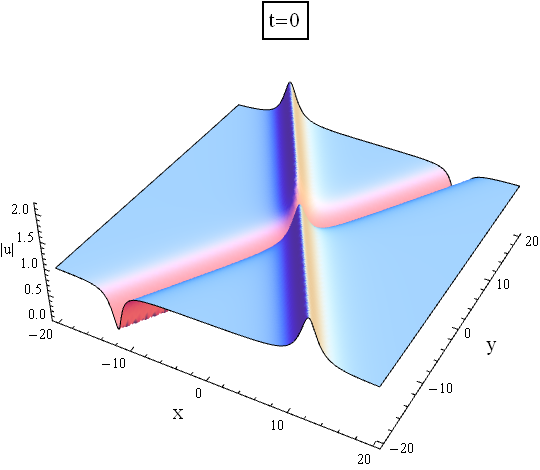}\hspace{2.0cm}
   \includegraphics[scale=0.255, bb=0 0 385 567]{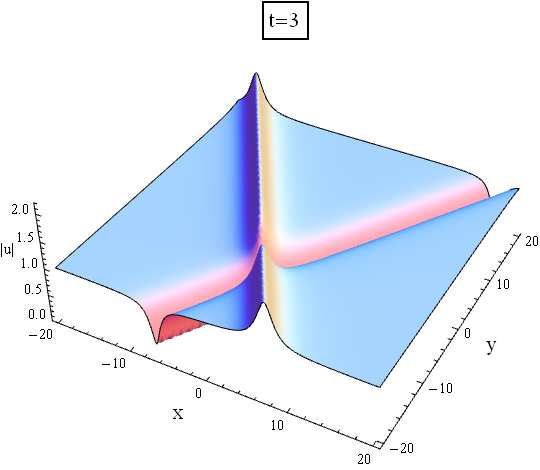}
   \caption{The interactions of  rational travelling wave solution for nonlocal DS-I equation at different time points, with parameters: $r_{1}=2$, $e_{1}=0$,  $f_{1}=1$.}
   \label{picture-label}
   \end{center}
\end{figure}

Although this solution is  generated from the 1-st iteration of DT, it contains two  rational travelling waves laying on  different trajectories.
For example, if one takes $r_{1}=2$, a time evolution process for this solution is displayed in Fig.3. When $t\rightarrow \pm\infty$, two  rational travelling waves move away from each other on a constant background, which behaviours like an interaction between a bright and dark  soliton.

Especially, when $r_{1}=1$, the above solution is reduced to:
\begin{eqnarray*}
u_{1}(x,t)=1+\frac{4 i (2 t-2 \mathit{e}_{1}+i)}{4 (\mathit{e}_{1}- t)^2-4 (x+i \mathit{f}_{1})^2+1}.
\end{eqnarray*}

This is an interesting one-dimension rational soliton solution for nonlocal NLS equation. Actually,
under the variable transform $u \rightarrow \tilde{u}=u e^{- i \alpha^2 t}$, then $\tilde{u}$ satisfy the nonlocal NLS equation which are reduced from nonlocal system by removing the  y-independence of the equation. Generally, utilizing this parameters choosing rules in nonlocal DS-I system, we may also derive multi-rogue waves  which are just nonlinear combinations of these fundamental patterns.

\subsection{Multi-rational solution  in partially $\cal{PT}$-symmetric  nonlocal DS-I equation}
Normally, N-rational solutions are generated from $N$ eigenfunctions with $4n$ parameters via Darboux transformation. With appropriate combinations of these parameters, it will present different dynamical  patterns, including singular multi-rogue waves blow-up in the finite time interval  and the  nonsingular mixture of fundamental rogue wave and rational travelling wave solutions.

For instance, taking $N=2$ in formula (\ref{potentialu1})-(\ref{potentialw1}) and choose the special parameters as:
$
\varphi _1=2 \pi,\  \varphi _2=2 \pi,\ r_2=1/r_1,\  \mathcal{F}_1=0,\ \mathcal{F}_2=0.$ It generates  the two-rogue wave solution with  particular singulary time points which are obtained by analysing its singularity. The imaginary part in the denominator is $ 16 xyt[  r_1^4 (r_1^4+1 )^2]$. Therefore, when $t=0$, the imaginary part of the denominator vanishes while the real part becomes
\begin{center}
$\Sigma_{s}(x,y)=\left[x^2 r_1^2 \left(r_1^2-1\right){}^2-y^2 r_1^2 \left(r_1^2+1\right)^2+3 r_14\right]^2+24 y^2 r_1^8-12 x^2 r_1^6 \left(r_1^4+1\right)$.
\end{center}
Obviously, this part will give rise to the singularities on above surface $\Sigma_{s}(x,y)=0$ at $t=0$. Next, if $y=0$,  the singularity time for this solution will happen on a finite interval $[t_{-}, t_{+}]$,\ where $t_{\pm}=\pm \frac{|r_1|^3 \sqrt{3 \left(r_1^2-1\right){}^2+4 r_1^2}}{\left|r_1^2-1 \right|\left(r_1^4+1\right)\sqrt{ \left(r_1^4-r_1^2+1\right)} }$.
Once $t$  falls into the interval, there will be two pairs of singularity points distribute centered on $x$-axis. And these points are substantially the real roots of a quartic equation dependent on variable $y$.
However, the number of the pairs down to one if $t$ is locate on the edges of the interval. At last, the real part of the denominator is proved to be definite positive if $x=0$.  As an example, when $r_1=2$, we find that the solution rises from a nearly constant background at $t=-\infty$, and then it appears a cross-shape wave in a intermediate time near $t=t_{-}$. However, finally this wave explodes to infinity at $t=t_{-}$. Once the solution exploded, the evolution of the wave will cease. There are also similar phenomena which appeared in the second-order and two-rogue waves of local DS-II equation\cite{YangDSII}. What is shown in fig.4 is the evolution of a two-rogue waves interaction together with the coming up of singularities. Especially, when $r_{1}=2$, the corresponding time interval is about $[-0.285287, 0.285287]$, so that the singular time $t=-0.2$ shown in fig.4 accurately falls into this interval.
\begin{figure}[htb]
   \begin{center}
   \includegraphics[scale=0.235, bb=0 0 385 567]{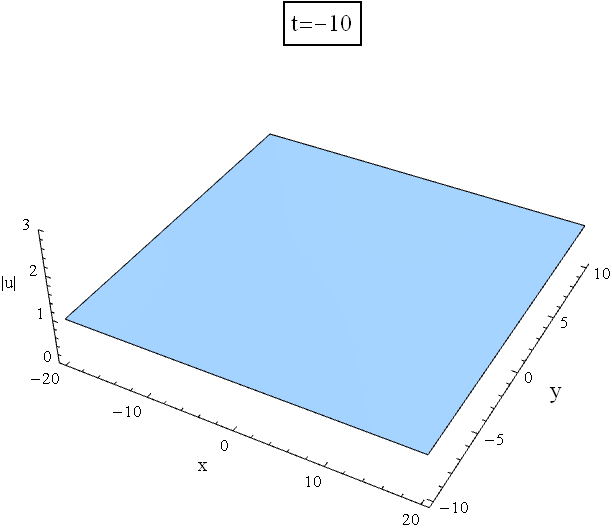}\hspace{2.35cm}
   \includegraphics[scale=0.235, bb=0 0 385 567]{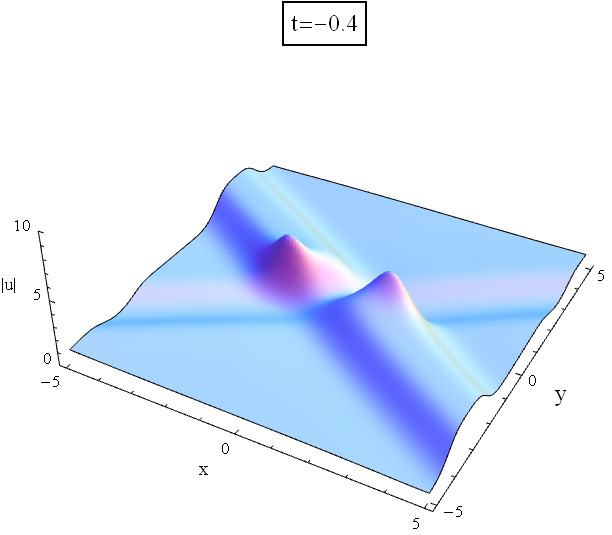}\hspace{2.35cm}
   \includegraphics[scale=0.235, bb=0 0 385 567]{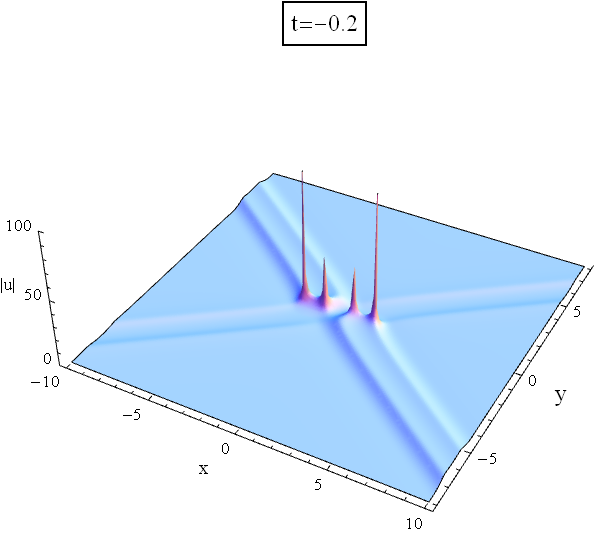}
   \caption{Time evolution of a two-rogue wave  solution for nonlocal DS-I equation with singularity points in pairs, with parameters:
   $\varphi _1=2 \pi,\  \varphi _2=2 \pi,\ r_1=2,\ r_2=1/r_1,\  \mathcal{F}_1=0,\ \mathcal{F}_2=0.$ }
   \label{picture-label}
   \end{center}
\end{figure}

Another novel hybrid multi-rogue wave pattern is obtained by  taking $N=2$ in formula (\ref{potentialu1})-(\ref{potentialw1}) with the parameters:
\ $
\varphi _1=\pi,\ r_1=1,\
\varphi _2=\frac{\pi }{2},\ r_2=1,\  \mathcal{F}_1=0,\ \mathcal{F}_2= \mathit{e_{2}}+ i \mathit{f_{2}}$, which leads a two-rational solution:
\begin{eqnarray}\label{hybirdRws}
u_{2}(x,y,t)=\frac{G(x,y,t)}{F(x,y,t)},
\end{eqnarray}
where,
\begin{eqnarray}
&&F(x,y,t)=4 y^2 \left(1+4(\mathit{e_{2}}-t)^2\right)-\left(16 y^2+ 16 t^2+4 \right) \left(x+i \mathit{f_{2}}\right)^2+\left( -4t^2 + 4t \mathit{e_{2}}+3\right)^2+4\mathit{e_{2}}^2,  \nonumber \\
&&G(x,y,t)=4[(2t-2i)^2+1]\left(x+i \mathit{f_{2}}\right)^2-[(2t-\mathit{e_{2}})^2+(i \mathit{e_{2}}-1)^2-4][(2t-\mathit{e_{2}})^2+(i\mathit{e_{2}}+3)^2-4]  \nonumber \\
&&\hspace{1.45cm}+4[4(i(\mathit{e_{2}}-t)+1)^2-1]y^2. \nonumber
\end{eqnarray}

With two free parameters given in expression (\ref{hybirdRws}),  it  can produces  an interesting novel hybrid pattern.  This pattern is described by the interaction of line rogue wave with dark and anti-dark travelling wave solution. In other words,  three different patterns appear at the same time in one solution. For any $\mathit{f_{2}}\neq0$, when $x=0$, the imaginary part of the denominator in solution  (\ref{hybirdRws}) becomes zero while the real part is  positive definite. Therefore, this solution is nonsingular except for $\mathit{f_{2}}=0$. Furthermore, with adequate combinations of   $\mathit{e_{2}}$ and $\mathit{f_{2}}$, solution  (\ref{hybirdRws}) can generate several interesting structures.

\begin{figure}[htb]
   \begin{center}
   \includegraphics[scale=0.225, bb=0 0 385 567]{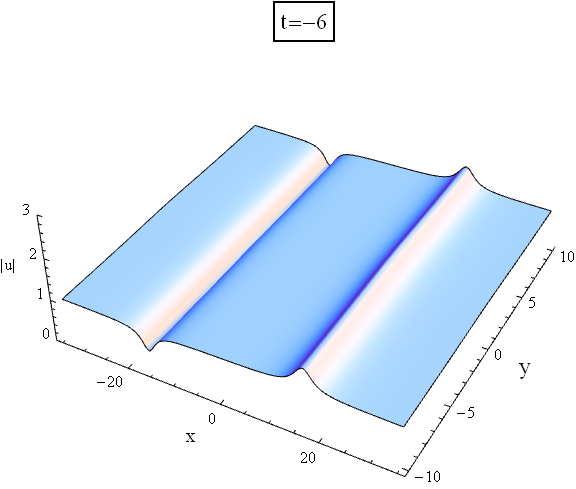}\hspace{2.5cm}
   \includegraphics[scale=0.225, bb=0 0 385 567]{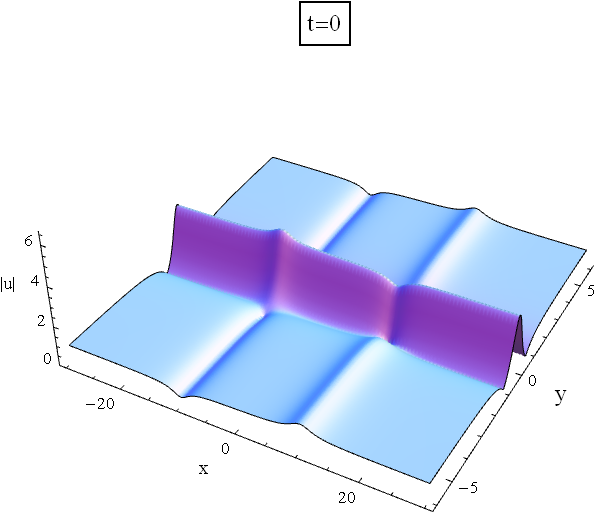}\hspace{5.0cm}
   \includegraphics[scale=0.225, bb=0 0 385 567]{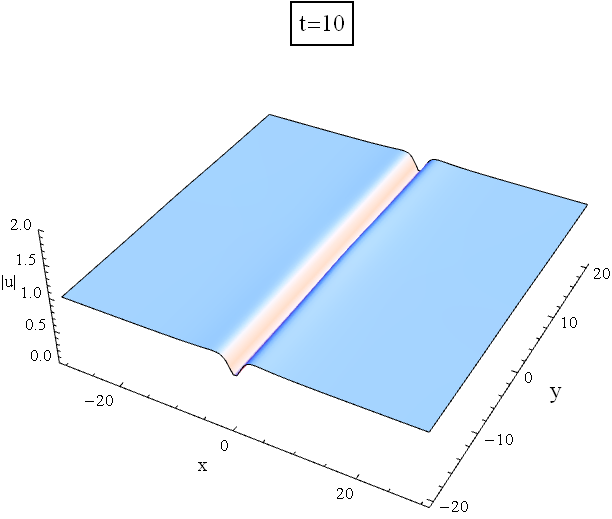}\hspace{2.5cm}
   \includegraphics[scale=0.225, bb=0 0 385 567]{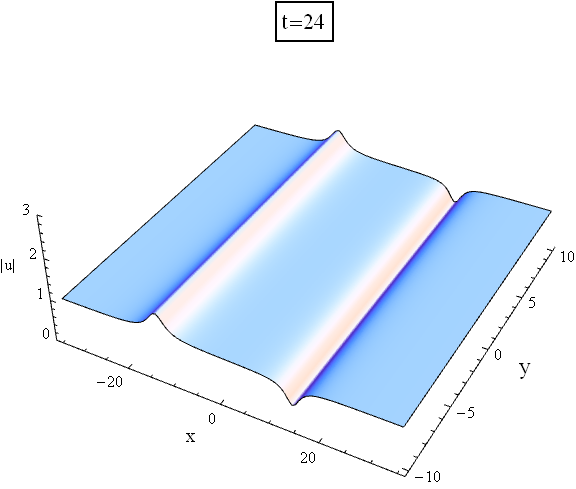}
   \caption{ The interactions of line rogue waves  with dark and anti-dark  rational travelling waves in the nonlocal DS-I equation,\ where the parameters are: \ $\varphi _1=\pi,\ r_1=1,\ \varphi _2=\frac{\pi }{2},\ r_2=1,\  \mathcal{F}_1=0,\ \mathcal{F}_2= 10+ 2 \textrm{i}$.}
   \end{center}
\end{figure}
For example,  choosing $\mathit{e_{2}}=0$  with $\mathit{f_{2}}=1$,  there is  a fundamental line rogue wave and a travelling wave interaction at about $t=0$. When $t\rightarrow\pm\infty$, it approaches two rational travelling waves  which slowly move away from each other.  Next, if one takes a larger value in $\mathit{f_{2}}$, i.e.,  $\mathit{e_{2}}=0$ and $\mathit{f_{2}}=10$.  This solution behaviours more like  a fundamental line rogue wave. This is because the amplitude of the travelling wave is much smaller than that of the rogue wave. However, as $t\rightarrow\pm\infty$, the rogue wave part decades very fast to a constant while the travelling wave portion continue moving apart.

Moreover, considering $\mathit{e_{2}}$ as a nonzero constant: $\mathit{e_{2}}=10$, $\mathit{f_{2}}=2$.  As it is shown in Fig.5 that when $t\rightarrow -\infty$, a dark and anti-dark rational travelling waves  move away from each other. It is also shown that a hybrid pattern of line rogue wave with dark and anti-dark travelling waves interactions appear around $t=0$. Afterwards, the line rogue wave soon disappears, then the dark and anti-dark rational travelling waves intersect and interact at about $t=10$, then they separate and move away from each other in an opposite direction as $t\rightarrow +\infty$.

\subsection{High-order rational solution in  the partially $\cal{PT}$-symmetric  nonlocal DS-I equation}

The high-order rational solution is another subclass of  rational solutions which exhibit different dynamics with multi-rational solution. And they can be obtained through the high-order Darboux transformation constructed in \emph{\textbf{Theorem }}1. Firstly, the second order rational solution is reduced from formula (\ref{GDTpotentialu}) by setting $N=1$.  Next, taking $\epsilon=1$, $\alpha=1$, $\varphi_{1}=\pi/4$, $r_1=1$, $\mathcal{F}_1=\mathit{e}_{1}$ for instance, here $\mathit{e}_{1}$ is a free real parameter, then solution $u_{1}^{[1]}(x,y,t)$  becomes
\begin{eqnarray}\label{highordersolution}
  u_{1}^{[1]}=-1+\frac{16 (1+2 i \mathit{e}_{1})[(-ix+y)^2+4t ]+16 i \left(2  x^2 + 2y^2+4 \mathit{e}_{1}^3+\mathit{e}\right)+24 \left(4 \mathit{e}_{1}^2+1\right)}{\left[8 t-2 (-ix+y)^2+4 \mathit{e}_{1}^2+3\right]^2+2 \left[4 \mathit{e}_{1}(-ix+y)-2 (-ix-y)\right]^2 + 8(-ix+y)^2+16 \mathit{e}_{1}^2}.
\end{eqnarray}

This solution is a very special case in the high-order solution for nonlocal DS-I equation. It is because solution (\ref{highordersolution}) has the same form with the second-order rogue wave solution in local DS-II system except for a simple variable transformation $x\rightarrow i x,\ t\rightarrow -t$. However, we make this transformation at the price of causing complex singularities to solution (\ref{highordersolution}). And these singularities are moving with the time. Furthermore, by another transformation $u_{1}^{[1]}(x,y,t)\rightarrow (-\sqrt{2}) u_{1}^{[1]}(ix,y,t-\frac{3}{8})$, solution (\ref{highordersolution}) becomes the second-order rogue wave solution for local DS-II equation which is derived in \cite{YangDSII} via the bilinear method.

Moreover, the nonsingular solutions can be also reduced from the second order rational solution by taking $\varphi_{1}=\pi/2$, $\mathit{e}_{1}=1$, and this yields a nonsingular high-order rational travelling wave solution but not a rogue wave.

\section{General rational solution  for  the partially $\cal{PT}$-symmetric  nonlocal DS-II equation}
In this section, as what we have shown for nonlocal DS-I system, we construct the general rogue wave solution for nonlocal DS-II equation and  analyze the dynamics of these rogue waves. In addition, we also exhibit other types of rational solutions which are reduced from the Darboux transformation.

\subsection{Fundamental rogue waves for nonlocal DS-II}
To derive the fundamental rogue waves for nonlocal DS-II equation, we first need to present the general one-rational solution of the first order, which is obtained by  taking $n=1$ in formula  (\ref{un+1})-(\ref{wn+1}):
\begin{eqnarray}
&&u_{1}(x,y,t)=1-\frac{2i \emph{G}(x,y,t)+1}{F(x,y,t)}, \label{RwuDSII}\\
&&w_{1}(x,y,t)=\epsilon+2  [\ln(F(x,y,t))]_{xx}, \label{RwwDSII}
\end{eqnarray}
where,
\begin{eqnarray}
&&F(x,y,t)=\emph{G}^2(x,y,t)+\emph{H}^2(x,y,t)+\frac{1}{4\cos^2\varphi_{1}},\  \emph{p}_{1}=\frac{ r_{1}+\epsilon r_{1}^{-1}}{2},\ \emph{q}_{1}=\frac{ r_{1}-\epsilon r_{1}^{-1}}{2},   \nonumber \\   \nonumber
&&\emph{G}(x,y,t)=i\emph{p}_{1} x \sin \varphi_{1} -  \emph{q}_{1} y \sin \varphi_{1} + (\emph{p}_{1}^{2}+  \emph{q}_{1}^{2}) t \cos 2\varphi_{1} + (e_{1}+\frac{1}{2}\tan \varphi_{1}),  \\
&&\nonumber \emph{H}(x,y,t)=i\emph{q}_{1}x \cos \varphi_{1} - \emph{p}_{1} y \cos \varphi_{1}  - 2 \emph{p}_{1} \emph{q}_{1} t \sin 2\varphi_{1}-( \mathit{f_{1}}+\frac{1}{2}).
\end{eqnarray}
For this rational solution,   different dynamics can be exhibited depending on the parameters  $\epsilon$, $\varphi_{1}$  and $r_{1}$. By performing solution analysis analogous to that in nonlocal DS-I equation, we find that:

(i).When $\epsilon=1$, $r_1=1$, then $|\lambda_{1}|=1$. In this case, we obtain the fundamental rogue wave solution.    And this rogue wave arises from a constant background as $t\rightarrow-\infty$ and develops finite-time singularity on a certain spatial location.  To be specific, when $\varphi_{1}=\frac{(2k-1)\pi}{2}$,  we have $u(x,y,t)=1$. For $\forall \varphi_{1}\neq \frac{(2k-1)\pi}{2}$, the  imaginary part of the denominator is\ $ x \sin \varphi_{1} \left(\tan\varphi_{1} + 2t \cos2\varphi_{1}  \right)$. If $x=0$, it can be shown the real part of the denominator is nonzero. Hence no singularities will appear in this situation. However, if $t=\frac{2 e \cos\varphi_{1} +\sin\varphi_{1}}{-2\cos2\varphi_{1} \cos\varphi_{1}}$,  the singularities will occurs at one time point and locates  on the certain elliptic curve in the $(x,y)$ plane:
\begin{eqnarray*}
(2y \cos^2\varphi_{1} + \cos \varphi_{1})^2-(x \sin2\varphi_{1})^2+1=0.
\end{eqnarray*}
For example, if we take $\varphi_{1}=-\pi/6$ with $e_{1}=0$, then the  singularity time occurs at  $t_{c}=\sqrt{3}/3$. Before this point, the rogue wave is  nonsingular and it shows some  cross-shape or interaction phenomena, which are quite different from the solution dynamics exhibited by the rogue wave in the local DS systems.  The dynamic evolution of this rogue wave is presented in Fig.6, including its shape at singular time $t_{c}$, where a truncation surface is obvious to see from the figure. It is remarkable that this kind of exploding fundamental rogue waves in the  partially $\cal{PT}$-symmetric  nonlocal sytems were first obtained in \cite{BoTransformations} by a simple ``transformation" method. Here, by using DT theory, one can easily generalize  solution formulas and parameters choices to the multi-rogue waves.
\begin{figure}[htb]
   \begin{center}
   \includegraphics[scale=0.280, bb=0 0 385 567]{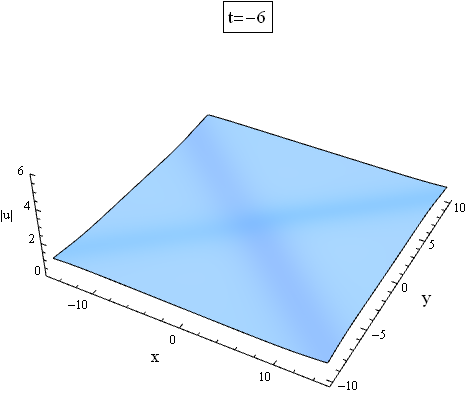}\hspace{1.5cm}
   \includegraphics[scale=0.280, bb=0 0 385 567]{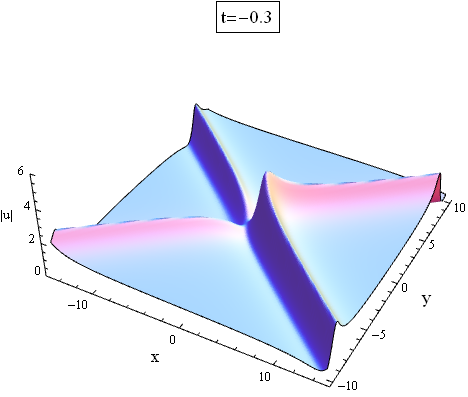}\hspace{1.5cm}
   \includegraphics[scale=0.280, bb=0 0 385 567]{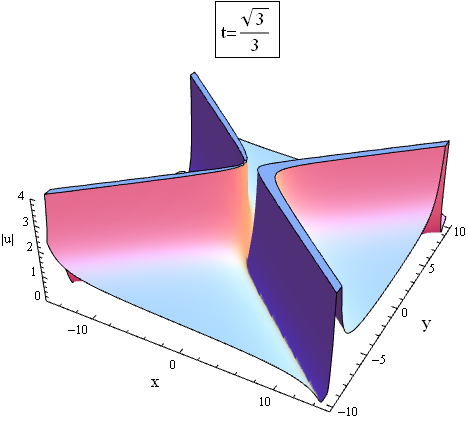}
   \caption{Fundamental cross-shape rogue wave in the the partially $\cal{PT}$-symmetric  nonlocal DS-II equation, behaviours from the constant background to the exploding  time, with parameters: $r_1=1$, $\varphi_{1}=-\pi/6$, $e_{1}=0$,  $f_{1}=0$. }
   \label{picture-label}
   \end{center}
\end{figure}

In addition, if $r_1=1$, $\varphi_{1}=k\pi$, as what we have found in the nonlocal DS-I system, this solution can  also degenerate into a (1+1)-dimensional line rogue wave solution:
\begin{eqnarray*}
u_{1}(y,t)=1-\frac{4(1+2i t)}{4 t^2+4 y^2+1}.
\end{eqnarray*}
The graphs of this rogue wave solution are qualitatively similar to those in Fig.2.

(ii). When $\epsilon=-1$, $r_1=1$, i.e., $|\lambda_{1}|=1$. In generic case, if we require $f_{1}\neq-1/2$,  one can derive a nonsingular solution with three parameters, which is nothing but the rational travelling wave for the nonlocal DS-II equation.
In fact, this solution can be seen as the corresponding  counterpart of travelling wave solution for nonlocal DS-I system.  Similarly, the ridge of this solution lays approximately on two lines with opposite slope, which is:
\begin{eqnarray*}
\emph{l}_{1}:\ 2 t \cos \varphi_{1} \cos 2 \varphi_{1}-y \sin 2 \varphi_{1} + 2x \cos^2 \varphi_{1}+\sin\varphi_{1}+e_{1}=0,\\
\emph{l}_{2}:\ 2 t \cos \varphi_{1} \cos 2 \varphi_{1}-y \sin 2 \varphi_{1} - 2x \cos^2 \varphi_{1}+\sin\varphi_{1}+e_{1}=0.
\end{eqnarray*}
It is easy to see that the  angle of $\emph{l}_{1}$ and $\emph{l}_{2}$ is $2\varphi_{1}$. And the solution are moving along these two center lines with the evolution of time.
\begin{figure}[htb]
   \begin{center}
   \includegraphics[scale=0.280, bb=0 0 385 567]{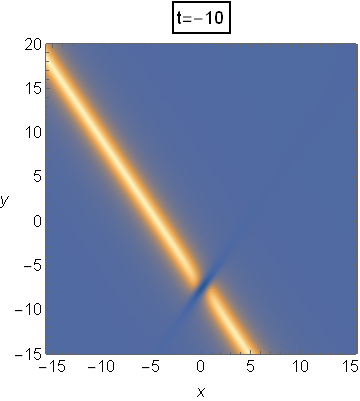}\hspace{0.6cm}
   \includegraphics[scale=0.280, bb=0 0 385 567]{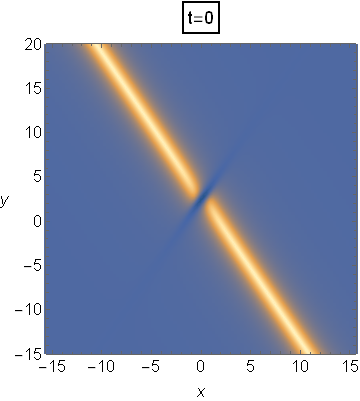}\hspace{0.6cm}
   \includegraphics[scale=0.280, bb=0 0 385 567]{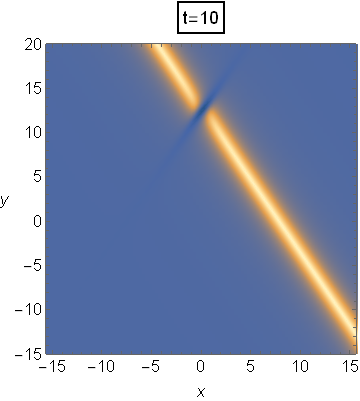}
   \caption{Rational travelling   waves interaction for the nonlocal DS-II equation, with parameters: $\epsilon =-1$, $\varphi_{1}=\pi/6$, $r_1=1,\ e_{1}=1,\ f_{1}=0$.}
   \label{picture-label}
   \end{center}
\end{figure}

Therefore, this lead us to make a summing up for the fundamental solutions of nonlocal DS systems. For the nonlocal DS-II equation, the value of $\epsilon$ determines the type of the fundamental rational solution under the unitary module reduction condition: $|\lambda_{1}|=1$, which is quite different from the condition we previously discussed  in the system of nonlocal DS-I equation, where the solution patterns are classified by the fact that whether $\lambda_{1}$ is a real or pure imaginary number. In the following, as what have already been shown in the nonlocal DS-I equation,  these parameters conditions can also applied for every $\lambda_{i}$  and it will produce several patterns of multi-rogue waves for the nonlocal DS-II system.

\subsection{Multi-rational solution for nonlocal DS-II equation}
To obtain the multi-rouge waves, one should make use of multi eigenfunctions with the form given in (\ref{Eigenfunctions}).  And this will bring more free parameters. However, we have noted that for $\forall k,j$, the denominator in the integration $\Omega(\varphi_{j},\phi_{k})$ has the term:
$r_k r_j\left( r_k +r_j e^{i \varphi_j+i \varphi_k}\right)^3$.

Therefore, one should choose parameters $r_k$ and $\varphi _k$ carefully because that may cause indeterminacy to the solution.  Here we set all nonzero $r_k  \in \mathbb{R}$, and parameters are limited to the condition:  $r_k r_j\left( r_k +r_j e^{i \varphi_j+i \varphi_k}\right)^3\neq0$. More specifically:\\
(1). If $e^{i \theta _j+i \theta _k}=1$, then $r_k+r_j\neq0$;\ \ (2). If $e^{i \theta _j+i \theta _k}=-1$, then $r_k-r_j\neq0$;\ \
(3). Once $r_k = r_j$, then $e^{i \theta _j+i \theta _k} \neq -1$.

These parameters can not be taken directly on the possible singular value points.  However,  the above  restrictions might be removed through a limiting process. For example,  taking $n=2$ in formula (\ref{un+1})-(\ref{wn+1}), it  generates a family of general two-rational solution for nonlocal DS-II equation.
Firstly, if we choose the parameters $\epsilon=1,\ \alpha=i,\ \varphi _1=2\pi,\ r_1=1,\ r_2=1,\  \mathcal{F}_1=0,\   \mathcal{F}_2=0$ in (\ref{un+1})-(\ref{wn+1}) and take the limit $\varphi _2 \rightarrow \frac{\pi }{2}$, then this two-rational solution reduce to the one-dimensional fundamental rogue wave solution
\begin{eqnarray*}
 u_{2}(y,t)=-1+\frac{4+8 i t}{4 t^2+4 y^2+1}.
\end{eqnarray*}

Next, when $\varphi_{2}$ continuously changes between 0 and $2\pi$ except for some particular values like 0, $\pi/2$, $\pi$ and $2\pi$.  A family of  rational solutions can be found with singularities existing on the corresponding time interval. However, usually it is a tedious process to determine the accurate interval values. Therefore, as a concrete example, we choose the special parameters  $\epsilon=1,\ \alpha=i,\ \varphi _1=2\pi,\ r_1=1,\ \varphi _2=\frac{\pi }{4},\ r_2=1,\  \mathcal{F}_1=0,\   \mathcal{F}_2=0$ for the convenience in the following analysis. Then it becomes the  two-rational solution with its singularity time $t_{0}$ occurs no more at one time point but on a finite time interval $I_{s}$. In this case, $t_{0}\in I_{s}=[0.326232,\ 0.628852]$, while these two end points are the  approximate values of the real roots satisfing the  following  quadratic equation:
\begin{center}
  $16 \sqrt{2} \mathit{c}^2  - 20 \mathit{c}^2  - 64 \sqrt{2} \mathit{c} + 88 \mathit{c}+52 \sqrt{2}-73=0$.
\end{center}

In fact, this equation come from analysing the possible singular points from the denominator of the solution.
First of all, the imaginary part of the  this denominator is:
\begin{eqnarray*}
4 x\left[4 \left(4 \sqrt{2}-5\right) y^2+\mathcal{P}_{1}(t)\right],\ \text{where}\ \
\mathcal{P}_{1}(t)=4 t \left(4 \sqrt{2} t-5 t-16 \sqrt{2}+22\right)+52 \sqrt{2}-73.
\end{eqnarray*}
If $x=0$, it is verified that the real part of the denominator is positive definite. Hence $x=0$ can not be the singularity point. Next, for $\left(4 \sqrt{2}-5\right) y^2\geq0$, if there exists  $t_{0}$ such that $\mathcal{P}_{1}(t_{0})\leq0$, one can obtain a point $y_{0}$ s.t $4 \left(4 \sqrt{2}-5\right) y_{0}^2+\mathcal{P}_{1}(t_{0})=0$, thus the imaginary part becomes zero. Subsequently, put points $(y_{0}, t_{0})$ to the real part,  and then we can also solve out a real point $x_{0}$  to makes the real part be zero. However, if $\mathcal{P}_{1}(t_{0})>0$, then the imaginary part is proved to be nonzero. Hence the solution has no singularity under this condition.

The dynamics for this solution are shown in Fig.7.  We can see that as $t\rightarrow \pm \infty$, this solution approaches a  ``$\emph{\textbf{X}}$ "-shape background wave with very small amplitude. While the solution could reach  very high  maximum amplitude near $t=0$. Moreover, when $t=0.4$, which belongs to the singular interval $I_{s}$, the solution exploding at this singular time point.
\begin{figure}[htb]
  \begin{center}
   \includegraphics[scale=0.280, bb=0 0 385 567]{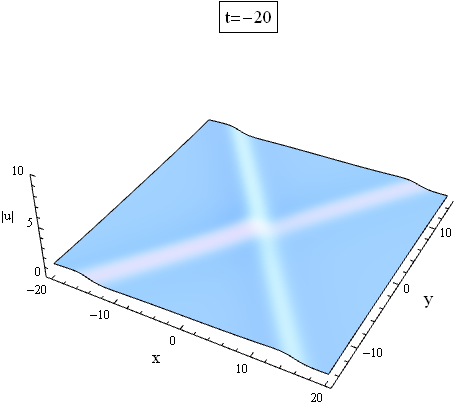}\hspace{1.5cm}
   \includegraphics[scale=0.280, bb=0 0 385 567]{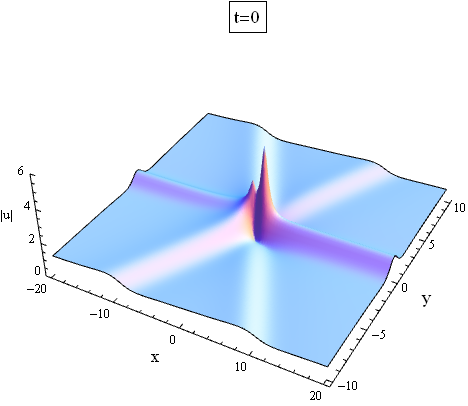}\hspace{1.5cm}
   \includegraphics[scale=0.280, bb=0 0 385 567]{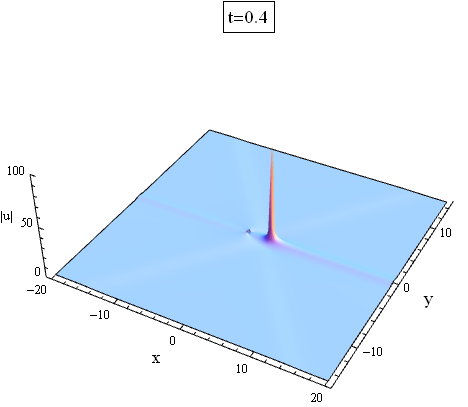}
   \caption{Dynamics of exploding two-rogue wave solution in the nonlocal DS-II equation, with parameters: $\epsilon=1,\ \alpha=i,\ \varphi _1=2\pi,\ r_1=1,\ \varphi _2=\frac{\pi }{4},\ r_2=1,\  \mathcal{F}_1=0,\   \mathcal{F}_2=0$. }
   \label{picture-label}
   \end{center}
\end{figure}

\subsection{High-order rational solutions for nonlocal DS-II}
As the case in the nonlocal DS-I quation,  the high-order rational solution for nonlocal DS-II equation can be constructed via the generalized binary DT (\ref{gDTDS-IIQN})-(\ref{gDTDS-IIwN}).
For  $N=2$, we have noted that the denominator in the integration $\Omega(\varphi_{j},\phi_{k})$  all contains the term: $1+e^{2 i \varphi _1}$. Thus, one should choose parameter carefully with $\varphi _1$.  For instance, choosing $\epsilon=1,\ \alpha=i$\ with $\varphi_{1}=2\pi,\ r_1=1,\ \mathcal{F}_1=0$, we obtain the second-order rational solution
\begin{eqnarray}
&&u_{1}^{[1]}=1+\frac{8 (1+2i t) [4i t (1+i t)+4i x-4 y^2+1]}{16 \left[ t^4+t^2 \left(-2i x+2y^2+1/2\right)+\left(ix+ y^2\right)^2\right]+8 i x+24 y^2+5},  \label{hiordersolutionDS-II} \\
&&w_{1}^{[1]}=-1+\frac{64 \left(-16 \left[ t^4+t^2 \left(-2 i x-6 y^2-3/2\right)+\left(ix + y^2\right)^2\right]-8 i x+8 y^2+3\right)}{\left(16 \left[t^4+t^2 \left(-2 i x + 2 y^2+1/2\right)+\left(ix+y^2\right)^2\right]+8 i x+24 y^2+5\right)^2}.
\end{eqnarray}
For this solution, it can be shown that it is singular for almost full time points except for a transient time interval.

Moreover, if we take the variable transformation $x\rightarrow -ix$,\ $t\rightarrow -t$, as what we have done for the high-order rational solution in nonlocal DS-I system,  then solution (\ref{hiordersolutionDS-II}) becomes
\begin{eqnarray}
 u_{1}^{[1]}=1+\frac{8 (1-2i t) \left[-4i  t (1- i t)+4 x-4  y^2+1\right]}{16 \left[ t^4+t^2 \left(-2x+2y^2+1/2\right)+\left(x+y^2\right)^2\right]+8 x+24 y^2+5}.
\end{eqnarray}
And this is the high-order rogue wave in the local DS-I equation which has been derived in \cite{YangDSI} through bilinear method. In this case, via a simple variable transformation, we derive this  well-posed high-order solution in the local DS-I equation from a ill-posed one with full-time singularity in the nonlocal DS-II equation.

\section{Summary and discussion}
In summary, we have derived general rogue waves in the partially $\cal{PT}$-symmetric  nonlocal DS-I and nonlocal DS-II equations. The tool we have used is the Darboux transformation method in soliton theory, and the solutions in these two equations are given in terms of determinants and  quasi-determinants, separately.
We have shown that the fundamental rogue waves in these two systems are rational solutions which arises from a constant background  and then  develops finite-time singularity on an entire hyperbola in the spatial plane at  the critical time (or at certain time interval, as what is shown in the de-focusing nonlocal DS-I equation).  We have also shown that multi rogue waves describes the interactions of several fundamental rogue waves. Especially,  a novel hybrid-pattern rogue wave is found, which contains three different types of waves in one solution. It exhibits different dynamics and is generated from the interaction of line rogue waves with dark and anti-dark rational travelling waves.    In addition,   some high-order travelling waves can be reduced from the high-order rational solutions,  and some singular solutions are also discovered, which can be transformed to the high-order rogue waves in the local DS systems through simple variables transformations.

Furthermore, it is interesting and meaningful  to compare these rogue wave in the nonlocal DS equations with those in the local DS equations (see refs.\cite{YangDSI,YangDSII}).  Firstly,  the parameter conditions for the generations of fundamental rogue waves are quite different between local and nonlocal DS equations. Secondly, we have known that for the local DS-II equation \cite{YangDSII}, rogue waves exist only when  $\epsilon=1$, but in the nonlocal DS-I equation, we have shown that rogue waves exist for both signs of nonlinearity  $\epsilon=\pm1$.
Thirdly, in the local DS equations, fundamental rogue waves are line rogue waves which are never blow up in finite time; While in the nonlocal DS equations, fundamental rogue waves have richer structures, including (1+2)-dimensional exploding rogue waves  and (1+1)-dimensional line rogue waves.   Although some non-generic multi-rogue waves and higher-order rogue waves of the local DS-II equation in ref.\cite{YangDSII} can also exploding in finite time, but the blowup only occurs at a single time point, unlike the fundamental rogue waves of the nonlocal DS equations where the blowup occurs on an entire hyperbola of the spatial plane.

Since partially $\cal{PT}$-symmetric physical systems has been shown possible applications in optics.  We expect these rogue-wave solutions could have interesting implications for the  partially $\cal{PT}$-symmetric   in multi-dimensions. Moreover, we hope these solutions could play a role in the physical understanding of rogue water waves in the ocean.

\section*{Acknowledgment}
The authors  like to show the deepest gratitude to Prof. Jianke Yang for his enthusiastic discussions and helpful suggestions.
This project is supported by the Global Change Research Program of China (No.2015CB953904), National Natural Science Foundation of China
(No.11675054 and 11435005), and Shanghai Collaborative Innovation Center of Trustworthy Software for Internet of Things (No. ZF1213).
The work of B.Y. is supported by a visiting-student scholarship from the Chinese Scholarship Council.

\end{document}